 \documentclass[final,3p,times]{elsarticle}

\usepackage{amssymb}
\usepackage{amsmath}

\usepackage{graphicx}
\usepackage{hyperref}
\usepackage{xcolor}
\usepackage{array}	

\journal{Acta Astronautica}

\begin{document}

\begin{frontmatter}



\title{Spacetime Contactability in Galactic Civilization Networks:\\ Contact Counts and Percolation}


\author[FRIS,TCU]{Kohji Tsumura\corref{cor1}} 
\cortext[cor1]{Corresponding author}
\ead{tsumura@astr.tohoku.ac.jp}

\affiliation[FRIS]{organization={Frontier Research Institute for Interdisciplinary Sciences, Tohoku University},
            addressline={6-3, Aramaki aza Aoba}, 
            city={Sendai},
            postcode={980-8578}, 
            state={Miyagi},
            country={Japan}}
            
\affiliation[TCU]{organization={Department of Natural Science, Faculty of Science and Engineering, Tokyo City University},
            addressline={1-28-1 Tamazutsumi}, 
            city={Setagaya},
            postcode={158-8557}, 
            state={Tokyo},
            country={Japan}}
            
\begin{abstract}
The number of Galactic civilizations does not determine whether they are in
contact. Allowing civilizations to appear at different times separates
listening, audibility, mutual-listening, and relay networks, with a distinct
reply-closing condition requiring a return signal within the sender's
lifetime. For each network we derive the contact counts and estimate the
connectivity threshold in a uniform population, then apply the results to a
structured Galactic disc, separating the detection range $d_{\rm max}$ from
the light horizon $cL$ through $\chi=d_{\rm max}/cL$. The contactability is
$\eta=\pi n d_{\rm max}^{2}$, where $n$ is the active civilization surface density.
Averaged over civilization appearances, the lifetime listening count is
$2\eta$ for all $\chi$ and, at fixed detection range, for any lifetime
distribution. Audibility and mutual-listening counts sum to $4\eta$, but
connectivity depends on threshold behavior rather than counts alone. As
$\chi\rightarrow0$, the listening threshold decreases from the coeval value
$4.51$ to $1.32\pm0.01$.
Assuming an ideal complete survey, our own non-detection gives the reference
bound $\eta(R_0)\le2.75$ at the Sun's Galactocentric radius. This benchmark
lies above the listening, audibility and relay thresholds throughout the
range studied. Unlike the
coeval model, the spacetime model can satisfy both this benchmark and a local
connectivity criterion, moving the local-contact frontier outward by
$3.2\,{\rm kpc}$. Under the same benchmark, civilizations hear at most
$5.5$ others on average, while the mean reply-closing count falls below unity
for $\chi\gtrsim1$. Contact may therefore be common even when completed
replies remain rare.
\end{abstract}

\begin{graphicalabstract}
\includegraphics[width=\linewidth]{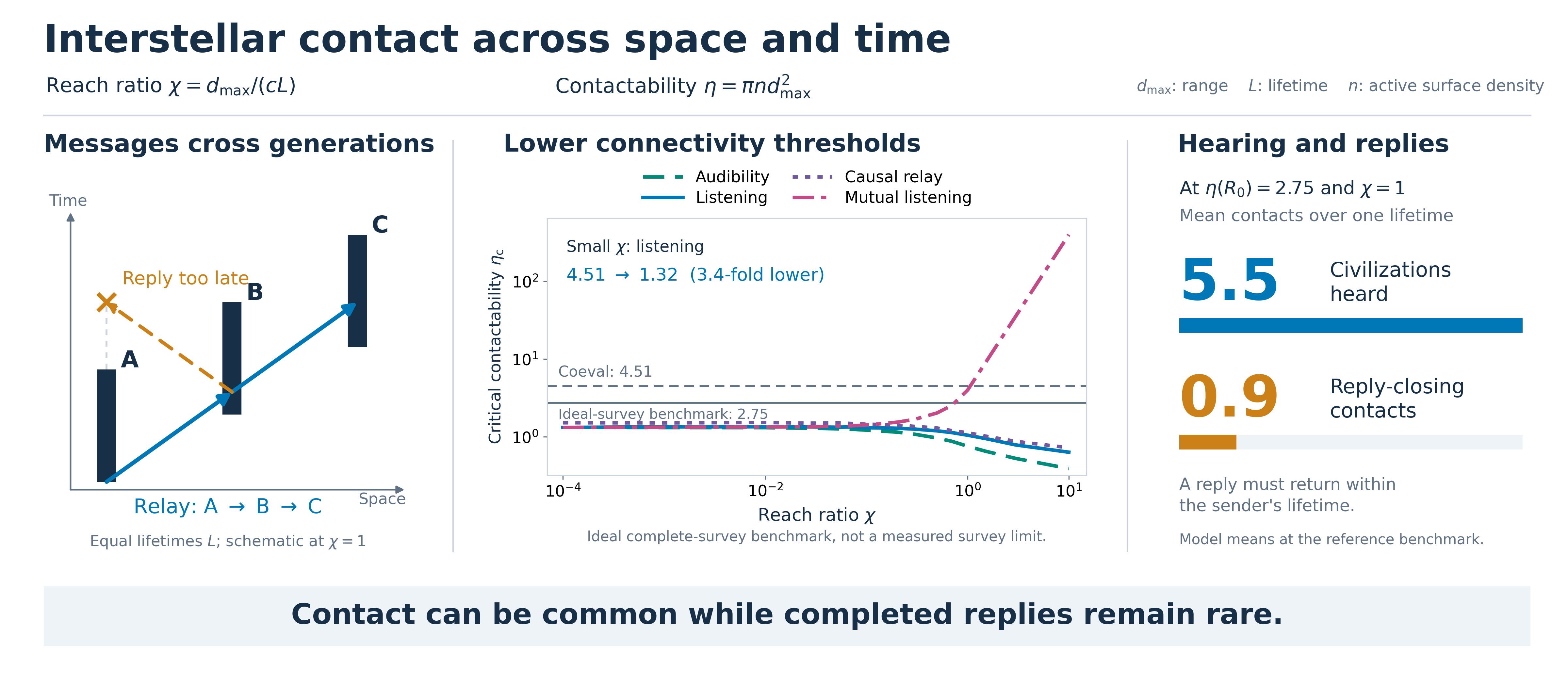}
\end{graphicalabstract}

\begin{highlights}
\item Finite lifetimes, offset start times and signal delays define four contact networks
\item Lifetime listening counts double at fixed active density and detection range
\item At small reach ratios, the listening threshold falls 3.4-fold from the coeval value
\item An ideal-survey benchmark and the local relay criterion can both be satisfied
\item Under the ideal-survey benchmark, large reach ratios allow contact but few replies
\end{highlights}

\begin{keyword}
SETI \sep interstellar communication \sep technosignature \sep Fermi paradox \sep temporal network \sep continuum percolation


\end{keyword}

\end{frontmatter}



\section{Introduction}
\label{sec:intro}

The Drake equation estimates how many communicative civilizations the Galaxy
contains \citep{Drake1965}, but not whether any of them are in contact. A
Galaxy may contain many civilizations that never
hear one another, or a few that form an interconnected network. For SETI,
detecting a signal and determining whether contact is common therefore require
different descriptions.

This paper addresses two questions. First, what does contact mean when
civilizations arise at different times and signals require time to travel?
Second, what does our own non-detection imply for Galactic contact? We
distinguish four relations: one-way listening, audibility in either direction,
mutual listening, and causal relaying. A separate reply-closing condition
specifies whether a two-way exchange can be completed. These relations define
four networks (Section~\ref{sec:three}), each with its own contact count and
connectivity threshold.

This paper is the third in a series that studies interstellar contact using
contact graphs. The central quantity is the \emph{contactability} $\eta$, the
expected number of civilizations within detection range of a given one
(Equation~\ref{eq:eta}). Paper~I \citep{GCNT1} established the framework for
a uniform population, and Paper~II \citep{GCNT2} extended it to a structured
Galactic disc. In both, civilizations capable of contact were assumed to
coexist. The present paper relaxes that assumption by
allowing signals to connect civilizations that need not be alive at the same
time. As a result, listening, audibility, mutual listening, and relaying
become distinct relations with different thresholds and network properties.
Paper~IV \citep{GCNT4} closes the series by letting the communicative
lifetime and the detection range differ from one civilization to another.

Throughout the paper we compare these thresholds with the benchmark of
Paper~II, $\eta(R_0)\le2.75$, which follows from our own non-detection under
the assumption of an ideal complete survey rather than from any actual
search (Section~\ref{sec:nondetection}). In the coeval
framework no single density satisfies both that benchmark and the local
connectivity criterion. We ask whether this changes once civilizations are
distributed through time.

\subsection{Contactability and two lengths treated as one}
\label{sec:onelength}

\begin{table*}
 \centering
 \caption{Observational scale of the reach ratio $\chi=d_{\rm max}/cL$.
Detection ranges $d_{\rm max}$ are given relative to the reference contact
length of Paper~II, $cL=6.1\times10^{3}$~ly (corresponding to
$\eta(R_0)=1$), assuming targeted observations. Rows marked S use the
detectability estimates of \citet{Sheikh2025}, while rows marked GM are
derived from the sensitivity thresholds of \citet{GrimaldiMarcy2018}; the
two sets are listed separately because they adopt different observing
assumptions. Reach ratios are computed from unrounded detection ranges,
whereas displayed $d_{\rm max}$ values are rounded. Blank entries repeat the
preceding value.}
 \label{tab:chiscale}
 \begin{tabular}{llrrc}
  \hline
  Emitter & Receiver & $d_{\rm max}$ & $\chi$ & Ref.\\
  \hline
  Voyager transmitter        & SKA1-Mid, 1~hr & $0.97$~ly  & $1.6\times10^{-4}$ & S  \\
  Mobile-telephony leakage   & SKA1-Mid, 1~hr & $4.0$~ly   & $6.6\times10^{-4}$ & S  \\
  Deep Space Network uplink  & SKA1-Mid, 1~hr & $65$~ly    & $0.011$            & S  \\
  Planetary radar            & SKA1-Mid, 1~hr & $12$~kly   & $1.97$             & S  \\
  \hline
  Planetary radar            & Past surveys   & $42$~ly    & $0.0069$           & GM \\
                             & Arecibo        & $1.17$~kly & $0.192$            & GM \\
                             & FAST           & $1.9$~kly  & $0.309$            & GM \\
                             & SKA2 (goal)    & $3.7$~kly  & $0.607$            & GM \\
  \hline
 \end{tabular}
\end{table*}

Paper~I identified two characteristic thresholds:
$\eta_{\rm contact}\simeq1$, above which a civilization typically has at least
one contact partner, and $\eta_{\rm network}$, above which a giant connected
component appears. The latter was measured as $4.33\pm0.13$ in finite systems
and associated with the continuum value $4.51$
\citep{DallChristensen2002,GCNT1,GCNT2}.

Papers~I and~II described contact using a single scale, the light horizon
$cL$, the distance travelled by light during a communicative lifetime $L$.
Civilizations were assumed to be connected whenever their separation was less
than this distance. This treatment implicitly identifies two physically
distinct quantities: the detection range $d_{\rm max}$, set by transmitter and
receiver performance, and the light horizon $cL$, set by longevity. It also
assumes that civilizations capable of contact coexist. The present paper
relaxes both assumptions and introduces the reach ratio
\begin{equation}
\chi \equiv \frac{d_{\rm max}}{cL},
\label{eq:chi}
\end{equation}
which compares detectability with longevity. For $\chi<1$ the detection range
is shorter than the light horizon; for $\chi>1$ it extends beyond the
distances that permit mutual listening or a completed reply. The reach ratio
is the principal control parameter of this study.

Throughout this paper, contactability is defined by the detection range,
\begin{equation}
\eta \equiv \pi n\, d_{\rm max}^{2},
\label{eq:eta}
\end{equation}
where $n$ is the surface density of civilizations active at a given epoch.
At $\chi=1$,
this expression reduces to the contactability parameter used in Papers~I and
II.

To place $\chi$ on an observational scale, we adopt the reference
normalization of Paper~II, for which $\eta(R_0)=1$ corresponds to
$cL\simeq6.1\times10^{3}$~ly. Published detectability estimates for
technosignatures span both $\chi\ll1$ and $\chi\gtrsim1$
(Table~\ref{tab:chiscale}; see \citealp{Sheikh2025,GrimaldiMarcy2018}).
\citet{Siemion2015} quote $100$~pc, or $\chi\simeq0.05$, for a one-minute
survey of every star in the primary beam. For the same transmitter, however,
targeted and all-sky searches can yield reach ratios differing by orders of
magnitude, since an all-sky search shortens the range by one to two orders of
magnitude \citep{Wandel2015}. The value of $\chi$ therefore depends as
strongly on how the search is conducted as on the sensitivity of the receiver
or the power of the transmitter. We therefore explore the
range $10^{-4}\le\chi\le10$. The shortest detection ranges listed in
Table~\ref{tab:chiscale} are smaller than the Galactic disc thickness and
would require a three-dimensional treatment before the two-dimensional results
presented here could be applied quantitatively.

\subsection{Temporal descriptions of the Galactic population}
\label{sec:temporal}

The importance of temporal structure in SETI has long been recognized.
Previous studies incorporated communicative lifetimes, civilization
appearance rates, and time-dependent histories into descriptions of the
Galactic population
\citep{Kreifeldt1971,Wallenhorst1981,Cirkovic2004,
CirkovicVukotic2008,Maccone2010,Smith2023}.
These works emphasize that the Drake equation constrains abundance but does
not by itself describe when civilizations exist.

The quantities governing temporal structure are also constrained by
astrophysics. Studies of planetary ages \citep{Lineweaver2001},
habitable-zone lifetimes \citep{Rushby2013,WolfToon2015} and biosphere
persistence \citep{Tsumura2020} suggest that communicative lifetime $L$
should be treated as an independent quantity with its own physical
constraints. If civilizations have appeared over a period much longer than
$L$, their total number may greatly exceed the number alive at any given
time \citep{FreemanLampton1975}.
We describe this distinction using the number of generations
$n_{\rm gen}$ (Section~\ref{sec:formulation}), defined as the ratio of the
total number of civilizations that have existed to the number coexisting at
a given time. Together, $n_{\rm gen}$ and $\chi$ provide the two dimensionless axes of the
problem.

\begin{table*}
 \centering
\caption{Contact relations and their associated networks.
Here $\Delta t=t_j-t_i$ is the appearance-time difference and $d$ the
spatial separation; all relations additionally require
$d<d_{\rm max}$. The mean degrees are derived in
Section~\ref{sec:contactability}; the small-$\chi$ and large-$\chi$
expressions join at $\chi=1$, except for reply-closing, whose branches join
at $\chi=1/2$.
A relation is path-dependent when reachability depends on message history
rather than only on the pair. The final row gives the most restrictive
pairwise criterion considered here.}
 \label{tab:relations}
 \small
 \begin{tabular}{llccccc}
  \hline
  Network & Condition & Eq. & Directed & Path-dep. & $\langle k\rangle/\eta$, small $\chi$ & $\langle k\rangle/\eta$, large $\chi$\\
  \hline
  Listening & $|\Delta t - d/c| < L$ & \eqref{eq:listen} & yes & no & \multicolumn{2}{c}{$2$} \\
  Audibility & $\bigl|\,|\Delta t| - d/c\,\bigr| < L$ & \eqref{eq:aud} & no & no & $2+4\chi/3$ & $4-2/(3\chi^{2})$ \\
  Mutual listening & $|\Delta t| + d/c < L$ & \eqref{eq:mutual} & no & no & $2-4\chi/3$ & $2/(3\chi^{2})$ \\
  Relay & $s_{j}+d_{jk}/c-L < t_{k} < s_{j}+d_{jk}/c+\ell_{j}$ & \eqref{eq:relayhop} & yes & yes & \multicolumn{2}{c}{$1+\ell/L$} \\
  \hline
  Reply-closing & $|\Delta t| + d/c < L$ and $d<cL/2$ & \eqref{eq:mutual} & no & no & $2-4\chi/3$ & $1/(3\chi^{2})$ \\
  \hline
 \end{tabular}
\end{table*}

\subsection{What detection models assume about the receiver}
\label{sec:receiver}

A separate group of SETI studies asks not how many civilizations exist, but
how many signals are detectable
\citep{Smith2009,Smith2023,ForganRice2010,Grimaldi2017,
Grimaldi2018,GrimaldiMarcy2018,Balbi2018}. These studies explicitly
incorporate signal propagation and causal structure, but they differ in the
assumptions made about the receiver. Some ask whether a signal is crossing
the observer at the present epoch
\citep{Grimaldi2017,Grimaldi2018,GrimaldiMarcy2018}; others count the
signals that have not yet passed, so that detection may occur now or at any
later time \citep{Balbi2018}. Others give the receiver a finite duration of
its own: it may appear while the signal is still passing \citep{Smith2009},
the two lifetimes may have to overlap enough for signals to reach both
participants \citep{ForganRice2010}; or a reply may have to return before the
sender's lifetime ends \citep{Smith2023}.
Table~\ref{tab:windows} sets the corresponding receiver conditions side by
side.

The present paper adopts a population perspective in which receivers are
civilizations with finite communicative lifetimes, treated in the same way as
senders. Pairwise causal-contact conditions of this type have been studied
previously \citep{ForganRice2010,Smith2023}, and network-level simulations
with random appearance times have also been explored
\citep{Lares2020,Forgan2019}. Our objective is different: we distinguish four
contact relations within a common spacetime framework, derive their mean
contact counts analytically, and measure the corresponding percolation
thresholds, including that for causal relaying.

The study most closely related to the present work is
\citet{Grimaldi2018}, who showed that detectable signals may originate from
civilizations that no longer exist. Here we extend that idea from the
perspective of a single observer to that of the entire Galactic population,
turning a detection problem into a network problem.
Once contacts are represented as graph edges, connectivity and percolation
become natural population-level questions. Contact counts are derived in
Sections~\ref{sec:formulation} and~\ref{sec:contactability}, and their
connectivity consequences are examined in
Sections~\ref{sec:percdef}--\ref{sec:ghz}.

\subsection{Four networks, not one}
\label{sec:three}

Papers~I and~II also gave civilizations finite lifetimes, but assumed that
those capable of contact were contemporaries. Under that assumption, and
with the single length $d_{\rm max}=cL$ those papers used, contact reduces to
the geometric condition $d<d_{\rm max}$, and the three
pairwise relations introduced below become equivalent. Causal relaying does
not: a relayed message still accumulates propagation delay along its path
(Section~\ref{sec:recovery}).

Once appearance times are allowed to differ, contact separates into distinct
causal relations. A civilization may \emph{listen} to another without being
heard in return, may \emph{relay} information through intermediate
civilizations, or may participate in \emph{mutual listening} when both
civilizations hear one another while active. These relations generate four
networks (Table~\ref{tab:relations}). The listening graph is directed. Its
undirected counterpart is the \emph{audibility graph}, while requiring
reciprocity gives the \emph{mutual-listening graph}. The \emph{relay graph}
is also directed but depends on communication paths rather than individual
pairs. A fifth network, the \emph{reply-closing graph} of
\citet{Smith2023}, is introduced in Section~\ref{sec:compare}; it coincides
with the mutual-listening graph for $\chi<1/2$ and becomes a proper subgraph
for $\chi>1/2$.

These networks need not share the same contact counts, connectivity
thresholds, or percolation properties. The distinction arises because a
civilization's signals remain detectable after that civilization has stopped
transmitting \citep{Grimaldi2018}. As a result, audibility does not
necessarily imply reciprocity. The reach ratio $\chi$ controls the degree of separation between
the networks: for $\chi\ll1$ the three pairwise networks (listening,
audibility and mutual listening) are nearly identical, whereas for $\chi\gg1$ their properties diverge substantially. The
relay network stands apart at every $\chi$. Section~\ref{sec:contactability}
quantifies this dependence.


\section{Formulation}
\label{sec:formulation}

The model represents civilizations as finite-lifetime events in two spatial
dimensions of the Galactic disc and one temporal dimension. It adopts five
assumptions:

\begin{description}
\item[A1] A civilization transmits and listens over the same interval, so the
receiver is another finite-lifetime civilization rather than a permanent
observatory.
\item[A2] All civilizations share the same communicative lifetime $L$.
\item[A3] Civilization appearances are uniform in time, yielding a stationary
population.
\item[A4] Every recipient retransmits, so reachability is defined by
possibility rather than motivation.
\item[A5] All civilizations share the same detection range $d_{\rm max}$ and
emit isotropically.
\end{description}

These assumptions play different roles: A1 and A2 fix the temporal window
for contact, A3 makes a single stationary window sufficient, A5 sets the
normalization through the common detection range that defines $\eta$, and A4
makes the relay counts well defined whatever a civilization's willingness to
forward a message. We revisit each where it first enters the analysis, and
Section~\ref{sec:assumptions} examines what relaxing it would change.

\subsection{Civilizations as spacetime events}
\label{sec:events}

A civilization is represented as a spacetime event of finite duration. The
$i$-th civilization is specified by a position $(x_i,y_i)$ in the Galactic
disc and an appearance time $t_i$. It transmits and receives during the
interval $[t_i,t_i+L]$ (A1), after which it plays no further role. All
civilizations share the same communicative lifetime $L$ (A2).

A disc of radius $R$ therefore contains $N=\pi R^2 n$ active civilizations.
Since each civilization remains active for a time $L$, the appearance rate
per unit area is
\begin{equation}
\lambda \equiv \frac{n}{L},
\label{eq:intensity}
\end{equation}
and the set of civilization events
$\{(x_i,y_i,t_i)\}$ forms a Poisson process in $(x,y,t)$ space. Appearance
times are assumed to be uniform over the Galactic history considered (A3),
yielding a stationary population. Until
Section~\ref{sec:ghz}, we further assume a spatially uniform intensity
$\lambda$; the structured Galactic disc is introduced in
Section~\ref{sec:ghz}.

The contactability,
\begin{equation}
\eta=\pi n d_{\rm max}^2,
\end{equation}
is the mean number of coexisting civilizations within detection range of a
given civilization. Because $n$ counts only civilizations active at a given
time, $\eta$ characterizes the contemporaneous population. Temporal effects
enter through the contact relations of Section~\ref{sec:links}, not through
$\eta$ itself.

\subsection{Two dimensionless axes}
\label{sec:axes}

Separating the detection range from the communicative lifetime introduces the
reach ratio $\chi$ (Equation~\ref{eq:chi}). Allowing civilizations to appear
at different times introduces a second dimensionless quantity,
\begin{equation}
n_{\rm gen}\equiv\frac{T_{\rm G}}{L},
\label{eq:ngen}
\end{equation}
which measures the number of communicative generations over the Galactic
history $T_{\rm G}$. A population containing $N$ contemporaneous
civilizations therefore contains $M=N\,n_{\rm gen}$ civilizations in total. 
For Galactic timescales and plausible communicative lifetimes,
$n_{\rm gen}\gg1$, implying that the total number of civilizations that have
ever existed can greatly exceed the number present at any given time.

An unbounded uniform system is thus characterized by three dimensionless
parameters, $(\eta,\chi,n_{\rm gen})$. The contactability $\eta$ sets the
density of contemporaneous civilizations, while the contact counts depend on
both $\eta$ and $\chi$. The parameter $n_{\rm gen}$ instead sets the temporal
extent of the population, and affects connectivity rather than the
counts (Sections~\ref{sec:links} and~\ref{sec:span}).
Finite realizations introduce one additional ratio, $R/d_{\rm max}$, which
controls finite-size and boundary effects but is a property of the system
being modeled rather than a parameter of the model itself.

\subsection{The link conditions}
\label{sec:links}

\begin{figure*}
\centering
\includegraphics[width=\textwidth]{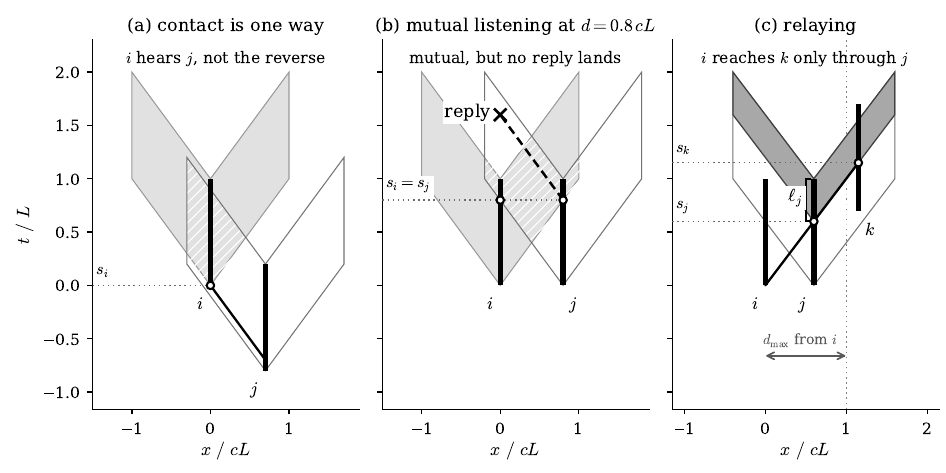}
\caption{
Spacetime representation of the three link conditions for $\chi=1$
($d_{\rm max}=cL$). One spatial coordinate is shown horizontally in units of
$cL$ and time vertically in units of $L$ ($c=1$). Civilizations are
represented by worldlines of duration $L$, and shaded regions indicate the
spacetime volume reachable by their transmissions within the detection range.
Open circles mark signal arrivals and brackets indicate the remaining
transmitting lifetime.
(a) Listening without reciprocity: $i$ receives a signal from $j$, but a
reply cannot reach $j$ before its communicative lifetime ends.
(b) Mutual listening: both civilizations receive each other's transmissions
while active. The dashed line illustrates a reply that arrives too late to
satisfy the reply-closing condition.
(c) Relaying: a message arriving at $j$ with residual lifetime $\ell_j$ is
retransmitted to a third civilization $k$. Although $k$ lies outside the
direct detection range of $i$, it remains reachable through the relay path.
}
\label{fig:spacetime}
\end{figure*}

The contact relations are illustrated in
Figure~\ref{fig:spacetime}. Consider two civilizations $i$ and $j$.
Their separation is $d_{ij}=\sqrt{(x_i-x_j)^2+(y_i-y_j)^2}$,
and their relative appearance time is
\begin{equation}
\Delta t_{ij}=t_j-t_i .
\label{eq:dt}
\end{equation}
All pairwise relations require $d_{ij}<d_{\rm max}$ and differ only in their
temporal condition.

The \emph{listening} relation holds when a signal emitted by $i$ overlaps the
listening interval of $j$,
\begin{equation}
\left|\Delta t_{ij}-d_{ij}/c\right|<L .
\label{eq:listen}
\end{equation}
Listening is directed. The \emph{audibility} relation holds when either
direction of Equation~\eqref{eq:listen} is satisfied,
\begin{equation}
\bigl||\Delta t_{ij}|-d_{ij}/c\bigr|<L ,
\label{eq:aud}
\end{equation}
while \emph{mutual listening} requires both directions,
\begin{equation}
|\Delta t_{ij}|<L-d_{ij}/c .
\label{eq:mutual}
\end{equation}
Mutual listening is therefore possible only for $d_{ij}<cL$, whereas
audibility extends to the full detection range $d_{\rm max}$. Because signals
can outlive their sources, audibility does not necessarily imply reciprocity;
a completed exchange requires the stricter reply-closing condition introduced
in Section~\ref{sec:compare}.

Relaying introduces a path-dependent relation. Let $s_j$ be the arrival time
at civilization $j$ of a message originating from $i$, and
\begin{equation}
\ell_j=t_j+L-s_j ,
\label{eq:residual}
\end{equation}
the transmitting lifetime remaining after arrival. A further civilization
$k$ receives the relayed message if
\begin{equation}
s_j+d_{jk}/c-L<t_k<s_j+d_{jk}/c+\ell_j ,
\label{eq:relayhop}
\end{equation}
with arrival time
\begin{equation}
s_k=\max\!\left(t_k,\;s_j+d_{jk}/c\right).
\label{eq:arrival}
\end{equation}
Each hop is limited by $d_{\rm max}$, but the distance from source to final
recipient is not. A chain of such hops is a time-respecting path
\citep{HolmeSaramaki2012}, and $s_k$ is an earliest-arrival time. Relaying
therefore carries a message to civilizations that the source could never have
reached directly.

The link conditions define a finite temporal reach. Listening and audibility
require
\begin{equation}
|\Delta t_{ij}|<(1+\chi)L ,
\label{eq:timereach}
\end{equation}
whereas mutual listening requires $|\Delta t_{ij}|<L$. The contact counts
derived in Section~\ref{sec:contactability} therefore become independent of
$n_{\rm gen}$ once the history is sufficiently long. Relay reachability is
different because relay chains may extend across many generations; its
convergence is examined in Section~\ref{sec:span}.

\subsection{Recovery of Papers I and II}
\label{sec:recovery}

For a uniform disc of radius $R$ containing $N$ coexisting civilizations,
\begin{equation}
\eta=N\left(\frac{d_{\rm max}}{R}\right)^2,
\label{eq:etauniform}
\end{equation}
which generalizes the contactability parameter of Papers~I and II by
replacing the light horizon $cL$ with the detection range $d_{\rm max}$.
Throughout Sections~\ref{sec:formulation}--\ref{sec:thresholds} we assume a
spatially uniform population in order to isolate temporal effects.
Section~\ref{sec:ghz} returns to the structured Galactic disc.

At $\chi=1$, the detection range and light horizon coincide,
$d_{\rm max}=cL$, and Equation~\eqref{eq:etauniform} reduces exactly to the
contactability parameter used in Papers~I and II. Combining
Equation~\eqref{eq:etauniform} with the Drake relation
\begin{equation}
N=DL,
\label{eq:drake}
\end{equation}
where $D$ collects the remaining Drake-equation factors
\citep{Drake1965}, gives $\eta=\chi^2 c^2DL^3/R^2$.
The scaling $\eta\propto L^3$ derived in Paper~I therefore applies only at
fixed $\chi$. At fixed detection range instead,
$\chi\propto L^{-1}$ and $\eta=DL\,d_{\rm max}^2/R^2$, 
so contactability depends linearly on communicative lifetime.

Imposing coevality further sets $\Delta t_{ij}=0$. The listening,
audibility, and mutual-listening conditions then all reduce to the geometric
criterion $d_{ij}<d_{\rm max}$, and the corresponding networks become the
static random geometric graph studied in Papers~I and II. Relay
reachability does not reduce in the same way because propagation delays
accumulate along a chain. Consequently, relay reachability remains a
strictly stronger condition even in the coeval limit.
The present framework is therefore a two-parameter extension of the earlier
theory, introducing the reach ratio $\chi$ and generation count
$n_{\rm gen}$ while retaining the original contactability parameter $\eta$.

As in Papers~I and II, Equation~\eqref{eq:etauniform} neglects boundary and
density-gradient corrections. These effects are unimportant for the uniform
medium considered here but are included in
Section~\ref{sec:ghz}.


\section{Spacetime contactability}
\label{sec:contactability}

Each contact relation defined in Section~\ref{sec:links} permits a range of
relative appearance times whose width depends on the separation $d$. Let
$W(d)$ denote this allowed time window. By Campbell's theorem
\citep{Chiu2013}, the corresponding mean degree for a Poisson process of
intensity $\lambda$ in $(x,y,t)$ is
\begin{equation}
\langle k \rangle =
\int_{0}^{d_{\rm max}}
\lambda\,W(d)\,2\pi d\,{\rm d}d
=
2\eta
\int_{0}^{1}
\frac{W}{L}\,
u\,{\rm d}u,
\label{eq:windowint}
\end{equation}
where $u=d/d_{\rm max}$. Since the window depends on distance only through $d/(cL)=\chi u$, every mean degree can be
written as $\langle k\rangle=\eta\,F(\chi)$, with a relation-dependent function $F(\chi)$.

\subsection{Listening}
\label{sec:listening}

For the listening relation, Equation~\eqref{eq:listen} permits an interval
of constant width $W(d)=2L$, independent of separation. Substituting into
Equation~\eqref{eq:windowint} gives
\begin{equation}
\langle k\rangle_{\rm lis}=2\eta.
\label{eq:klis}
\end{equation}
Equation~\eqref{eq:klis} counts the civilizations that hear a given one, so
for the directed listening graph it is the mean out-degree. Stationarity
implies the same mean in-degree. Unlike audibility and mutual listening, the listening count is
independent of the reach ratio $\chi$; throughout this paper, variations in
$\chi$ are considered at fixed $\eta$.

The result differs from those of Papers~I and II, where all civilizations
were contemporaries and $\langle k\rangle=\eta$. Distributing the same
lifetimes through time doubles the count. One contribution comes from
civilizations that appear while the signal shell passes their location,
yielding the swept-volume count $\eta$. A second contribution comes from
civilizations already listening when the signal arrives, contributing another
$\eta$.

Equation~\eqref{eq:klis} remains valid for an arbitrary lifetime
distribution, provided communicative lifetime and $d_{\rm max}^2$ are
uncorrelated. Assumption A2 is therefore stronger than required. The more
general case, including correlations between lifetime and detection range, is
considered in Paper~IV.

\subsection{Audibility and mutual listening}
\label{sec:audmut}

Audibility and mutual listening combine the listening relation with its
reverse. Audibility requires that at least one civilization hear the other,
whereas mutual listening requires both directions.
The corresponding time windows are
\begin{equation}
W_{\rm aud}(d)
=
2L+2\min\!\left(L,\frac{d}{c}\right),
\qquad
W_{\rm mut}(d)
=
2\left(L-\frac{d}{c}\right)_+ ,
\qquad
(x)_+\equiv\max(x,0),
\label{eq:windows}
\end{equation}
where the two listening windows overlap only for $d<cL$.

\begin{figure*}
\centering
\includegraphics[width=\textwidth]{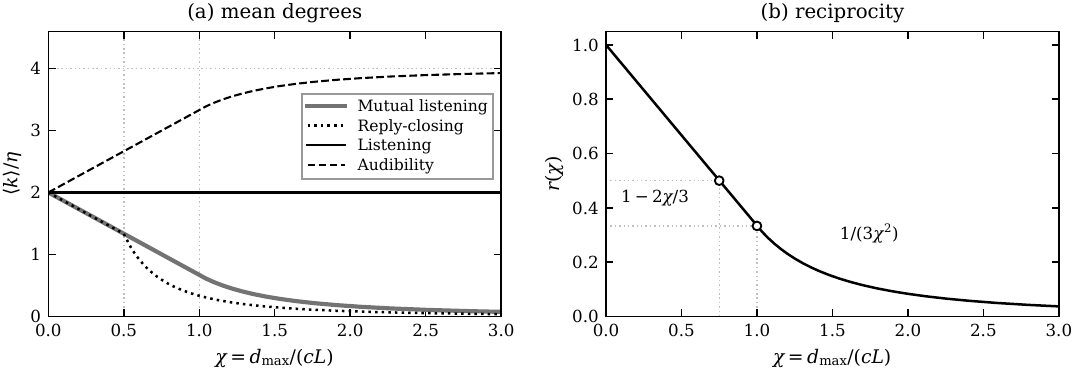}
\caption{
(a) Mean degrees, normalized by $\eta$, of the listening, audibility,
mutual-listening, and reply-closing networks as functions of $\chi$.
Dotted lines indicate the branch points ($\chi=1/2$, $1$), the common
small-$\chi$ limit ($2$), and the audibility saturation value ($4$). The
mutual-listening and reply-closing curves coincide for $\chi\le1/2$.
(b) Reciprocity $r(\chi)$ (Equation~\eqref{eq:recip}). Open circles mark
$(3/4,1/2)$ and $(1,1/3)$.
}
\label{fig:degrees}
\end{figure*}

Substituting Equation~\eqref{eq:windows} into
Equation~\eqref{eq:windowint} gives
\begin{equation}
\langle k\rangle_{\rm aud}
=
\begin{cases}
\eta\left(2+\frac{4}{3}\chi\right),
& \chi\le1, \\[4pt]
\eta\left(4-\frac{2}{3}\chi^{-2}\right),
& \chi\ge1 ,
\end{cases}
\label{eq:kaud}
\end{equation}
and
\begin{equation}
\langle k\rangle_{\rm mut}
=
\begin{cases}
\eta\left(2-\frac{4}{3}\chi\right),
& \chi\le1, \\[4pt]
\frac{2}{3}\eta\chi^{-2},
& \chi\ge1 .
\end{cases}
\label{eq:kmut}
\end{equation}
The two branches join continuously at $\chi=1$
(Figure~\ref{fig:degrees}). As $\chi\rightarrow0$, both counts approach
$2\eta$. As $\chi\rightarrow\infty$, audibility approaches $4\eta$ whereas
mutual listening decreases as $\chi^{-2}$. At fixed $\eta$, larger reach ratios therefore increase the audibility count
but reduce the fraction of listening links that are reciprocal.

The two counts obey the identity
\begin{equation}
\langle k\rangle_{\rm aud}
+
\langle k\rangle_{\rm mut}
=
2\langle k\rangle_{\rm lis}
=
4\eta ,
\label{eq:sumrule}
\end{equation}
which holds for all $\chi$ and, because it follows directly from the
listening count, for arbitrary lifetime distributions whenever
Equation~\eqref{eq:klis} does.
Lifetime distributions therefore affect only how the fixed total $4\eta$ is
partitioned between audibility and mutual listening. Variability in lifetime
can increase the mutual-listening count by allowing some pairs to satisfy the
reciprocity condition that would otherwise fail; by
Equation~\eqref{eq:sumrule}, the audibility count decreases by the same
amount.

\subsection{Reciprocity}
\label{sec:reciprocity}

The listening relation is directed. A natural measure of its asymmetry is
the reciprocity, defined as the fraction of directed links whose reverse link
also exists \citep{GarlaschelliLoffredo2004}. Because the reciprocated links
are precisely the mutual-listening pairs,
\begin{equation}
r \equiv
\frac{\langle k\rangle_{\rm mut}}
     {\langle k\rangle_{\rm lis}}
=
\begin{cases}
1-\tfrac{2}{3}\chi ,
& \chi\le1 , \\[4pt]
\tfrac{1}{3}\chi^{-2} ,
& \chi\ge1 ,
\end{cases}
\label{eq:recip}
\end{equation}
where the factor $\eta$ cancels. Reciprocity therefore depends only on the
reach ratio $\chi$ and not on civilization density
(Figure~\ref{fig:degrees}b).

The reciprocity provides a compact representation of the contact counts.
From Equations~\eqref{eq:recip} and~\eqref{eq:klis},
\begin{equation}
\langle k\rangle_{\rm mut}=2\eta r ,
\label{eq:mutviar}
\end{equation}
while the sum rule of Equation~\eqref{eq:sumrule} gives
\begin{equation}
\langle k\rangle_{\rm aud}=2\eta(2-r) .
\label{eq:audviar}
\end{equation}
Thus $\eta$ sets the overall level of contact, whereas $r(\chi)$ determines
the fraction that is reciprocated. In the coeval framework of Papers~I and
II, reciprocity is identically unity; in the spacetime framework it becomes
a geometric function of $\chi$.

\subsection{The relaying window}
\label{sec:relaywindow}

Relaying differs from listening and mutual listening because it is a
path-dependent rather than a pairwise relation. A civilization can forward a
message only during the remaining part of its communicative lifetime, given
by the residual lifetime $\ell_j$ of Equation~\eqref{eq:residual}.
Equation~\eqref{eq:relayhop} defines a reception window of length
$L+\ell_j$. The expected number of civilizations reachable by a relay is
\begin{equation}
\langle k\rangle_{\rm out}(\ell)
=
\eta\left(1+\frac{\ell}{L}\right).
\label{eq:kout}
\end{equation}
For the original source, $\ell=L$ and
Equation~\eqref{eq:kout} reduces to the listening count $2\eta$. As
$\ell\rightarrow0$, the expected relay count approaches $\eta$. Like the
listening count, Equation~\eqref{eq:kout} is independent of the reach ratio $\chi$.

Residual lifetimes do not decrease monotonically along a relay
chain. A civilization that receives the message after it has appeared can
forward it only for the part of its lifetime that remains, whereas one for
which the message is already arriving when it appears receives it at birth
and can forward it for the whole of $L$. Newly born civilizations therefore
restore the transmission time that propagation delays take away, so a long
chain of relays is possible even though every civilization is short-lived.

\providecommand{\brtwo}[2]{\begin{tabular}[t]{@{}l@{}}#1\\#2\end{tabular}}
\begin{table*}
 \centering
\caption{Receiver assumptions and their corresponding mean degrees
$\langle k\rangle$. All values are evaluated using the common
two-dimensional disc geometry of Papers~I and II for direct comparison.
A dash in the final column denotes a quantity included only for
comparison with the literature.}
 \label{tab:windows}
 \small
 \begin{tabular}{>{\raggedright\arraybackslash}p{2.4cm}llccl}
  \hline
  Assumption & Assumed by & Condition & $W(d)$ & $\langle k \rangle$ & Eq.\\
  \hline
  Coeval, instantaneous
    & Papers~I, II & $t_{i} = t_{j}$ & --- & $\eta$ & --- \\
  Observes at one epoch
    & \citet{Grimaldi2018} & $0 < t_{\rm obs} - t_{i} - d/c < L$ & $L$ & $\eta$
    & --- \\
  Born during the passage
    & \citet{Smith2009} & $0 < \Delta t - d/c < L$ & $L$ & $\eta$ & --- \\
  Both receive while active
    & \citet{ForganRice2010} & $|\Delta t| < L - d/c$ & $2(L - d/c)_{+}$
    & \brtwo{$\eta(2 - 4\chi/3)$, $\chi \le 1$}{$2\eta/(3\chi^{2})$, $\chi \ge 1$}
    & \eqref{eq:kmut} \\
  A reply reaches the sender
    & \citet{Smith2023} & \brtwo{$|\Delta t| < L - d/c$}{and $d < cL/2$}
    & $2(L - d/c)\,\mathbf{1}_{\{d < cL/2\}}$
    & \brtwo{$\eta(2 - 4\chi/3)$, $\chi \le \frac{1}{2}$}%
            {$\eta/(3\chi^{2})$, $\chi \ge \frac{1}{2}$}
    & \eqref{eq:kreply} \\
  \hline
  Same lifetime, listening
    & This paper & $|\Delta t - d/c| < L$ & $2L$ & $2\eta$
    & \eqref{eq:klis} \\
  Same lifetime, audibility
    & This paper & $\bigl|\,|\Delta t| - d/c \,\bigr| < L$ & $2L + 2\min(L,d/c)$
    & \brtwo{$\eta(2 + 4\chi/3)$, $\chi \le 1$}%
            {$\eta\left(4 - 2/(3\chi^{2})\right)$, $\chi \ge 1$}
    & \eqref{eq:kaud} \\
  \hline
 \end{tabular}
\end{table*}

\subsection{Comparison with the detection models}
\label{sec:compare}

Each receiver assumption of Section~\ref{sec:receiver} corresponds to a time
window $W(d)$, from which the contact count follows through
Equation~\eqref{eq:windowint}. Table~\ref{tab:windows} collects the results,
all recomputed in the two-dimensional geometry of Papers~I and II. Several
physically distinct assumptions give the same count,
$\langle k\rangle=\eta$: they differ in what is required of the receiver, not
in the number of detectable contacts.

The single-epoch observer needs particular attention. A signal is detected
whenever its shell crosses the observer, giving a window of duration $L$ and 
$\langle k\rangle_{\rm obs}=\eta$, independent of $\chi$. This is the quantity estimated by
\citet{Grimaldi2018} and \citet{GrimaldiMarcy2018}, and the quantity
constrained in the coeval framework of Papers~I and II. In contrast, treating
the receiver as a civilization with lifetime $L$ gives the lifetime-integrated
listening count $\langle k\rangle_{\rm lis}=2\eta$.
The difference arises entirely from the receiver assumption.

A stricter criterion is the reply-closing condition of \citet{Smith2023},
which requires a successful return signal in addition to mutual reception.
The corresponding count is
\begin{equation}
\langle k\rangle_{\rm reply}
=
\begin{cases}
\eta\left(2-\tfrac{4}{3}\chi\right),
& \chi\le\tfrac{1}{2}, \\[4pt]
\tfrac{1}{3}\eta\chi^{-2},
& \chi\ge\tfrac{1}{2},
\end{cases}
\label{eq:kreply}
\end{equation}
with the two branches joining continuously at $4\eta/3$. The reply-closing
graph coincides with the mutual-listening graph for $\chi<1/2$ and becomes a
proper subgraph for larger reach ratios. For $\chi>1$, $\langle k\rangle_{\rm reply} = \langle k\rangle_{\rm mut} /2$.

\section{Percolation in a spacetime network}
\label{sec:percdef}

Section~\ref{sec:contactability} established the local contact statistics of
the spacetime model. They do not determine connectivity, which depends on
paths rather than on neighbourhood size alone. In spacetime networks, three
features prevent a direct correspondence between count and connectivity.

\emph{First, the geometry is anisotropic.}
Space is isotropic but time is ordered, so the connection regions defined by
Equations~(\ref{eq:listen})--(\ref{eq:mutual}) are spacetime bicones and
their intersections rather than balls. Percolation thresholds therefore
depend on the shape of those regions as well as on the mean degree.

\emph{Second, relay reachability is path-dependent.}
Equation~\eqref{eq:relayhop} depends on the route by which a message arrives,
so a set of valid pairwise links does not by itself fix the reachable set.
Relay connectivity therefore cannot be read off the contact counts.

\emph{Third, connectivity can depend on history length.}
The pairwise contact counts become independent of $n_{\rm gen}$ once the
history is sufficiently long, whereas relay chains may continue to extend
across further generations. The counts
alone do not decide between these two behaviours.

The remainder of this paper addresses these issues through Monte Carlo
simulations. This section defines percolation in the four networks and the
associated measurements. Section~\ref{sec:thresholds} determines the
thresholds in a uniform medium, and Section~\ref{sec:ghz} applies the results
to the Galactic disc.

\subsection{What percolates in each network}
\label{sec:whatpercolates}

The four networks of Section~\ref{sec:three} are not equivalent, so
percolation must be defined separately for each.
Audibility, mutual listening, and reply-closing are undirected relations.
Their graphs are therefore random geometric graphs in three-dimensional
spacetime, and percolation refers to the appearance of a giant component:
one whose size stays proportional to the population as the system is made
larger, rather than shrinking to a vanishing share of it.

Listening is directed. Each civilization has both an out-component, composed
of civilizations reachable from it, and an in-component, composed of
civilizations that can reach it. Because time reversal exchanges these
components while leaving the ensemble unchanged, they percolate at the same
threshold. Throughout this paper we measure the out-component: the set of
civilizations a given one can reach, not a set within which every member can
reach every other.

Relay is fundamentally path-dependent. Reachability depends on transmission
histories through Equation~\eqref{eq:relayhop} and cannot be inferred from
pairwise links alone. Relay reachability is evaluated by propagating
earliest-arrival times. Since relay hops satisfy $\ell_j\le L$ and
$s_j\ge t_j$, every relay-reachable set is contained within the corresponding
listening out-component, implying $\eta_c^{\rm lis} \le \eta_c^{\rm relay}$.

\subsection{Monte Carlo construction and finite-size scaling}
\label{sec:mcmethod}

Civilizations are generated as a Poisson process in a finite spacetime box,
with distances measured in units of $d_{\rm max}$ and times in units of $L$.
The corresponding light speed is therefore $\chi^{-1}$ in simulation units.
Periodic boundaries are applied in the spatial directions and, for
undirected networks, in time. Listening and relay networks retain open
temporal boundaries because of causality.

The two kinds of network need different estimators. For the undirected
networks we use the Newman--Ziff procedure
\citep{NewmanZiff2000} and locate the threshold by the appearance of a
wrapping cluster. For the directed networks, wrapping in time is unphysical;
we instead propagate the reachable set forward from the first generation and
locate the threshold from the way its survival time scales with box size, as
described below and in \ref{app:thresholds}. Relay reachability is treated
as an earliest-arrival problem and evaluated using Dijkstra's algorithm
\citep{Dijkstra1959} with Equation~\eqref{eq:arrival} as the relaxation
rule.

Finite-size thresholds are extrapolated using
\begin{equation}
\eta_{\rm c}(m)
=
\eta_{\rm c}
+
a\,m^{-1/\nu},
\label{eq:fss}
\end{equation}
where $m$ is the box size in units of $d_{\rm max}$. For the undirected
networks we adopt the three-dimensional isotropic-percolation exponent
$1/\nu=1.1410(15)$ \citep{Wang2013}.
The relay transition is anisotropic and does not permit a reliable
fixed-aspect-ratio extrapolation over the sizes explored here. The directed
thresholds are therefore taken from the scaling of the median survival time
$t_{50}$ of the reachable set. The threshold is identified by consistency
with a dynamic exponent in the range $1.70\le z\le2.00$ for relay and with
the bracketing values $z=1$ and $z=2$ for listening. The resulting
thresholds are insensitive to the precise choice of $z$ within these
intervals. No universality class is assumed or determined.

The same code and estimator were applied to three problems with known
thresholds: the three-dimensional unit ball, the undirected
$\chi\rightarrow0$ limit of \citet{Hyytia2012}, and the causal counterpart of
\citet{HyytiaOtt2013}. The first two differ from the reference values by less
than $0.2$ per cent, and the causal benchmark agrees within the quoted
uncertainty (\ref{app:thresholds}). The numerical results reported below were accepted
only after these checks, the internal consistency tests of
\ref{app:thresholds}, and agreement with the analytic contact counts of
Section~\ref{sec:contactability} wherever those are available.


\section{Percolation thresholds in a uniform medium}
\label{sec:thresholds}

Connectivity appears when a giant component forms. The corresponding
critical contactability is denoted by $\eta_{\rm c}$. This section
determines $\eta_{\rm c}$ for the four contact networks introduced in
Section~\ref{sec:three}, across five decades in the reach ratio $\chi$.
For mutual listening at $\chi\ge1$, the $\chi^{2}$ scaling is derived
analytically and only its normalization is measured
(Section~\ref{sec:excluded}).

Papers~I and II considered the coeval limit, for which the network
threshold is $\eta_{\rm c}\simeq4.5$
\citep{DallChristensen2002,GCNT1,GCNT2}. Here we determine how that
threshold changes when civilizations are distributed through time and the
different contact relations are distinguished.
Our aim is to estimate thresholds for an infinite uniform medium.
The structured Galactic disc is considered separately in
Section~\ref{sec:ghz}.

\begin{table*}
 \centering
\caption{Critical contactabilities $\eta_{\rm c}$ of the four networks as a
function of the reach ratio $\chi$. Parameters $\eta$, $\chi$, and
$r(\chi)$ are defined by Equations~(\ref{eq:eta}), (\ref{eq:chi}),
and~(\ref{eq:recip}). Digits in parentheses denote uncertainties in the
last quoted digit: statistical for the audibility and mutual-listening
columns, and numerical for the listening and relay columns
(\ref{app:thresholds}).
}
 \label{tab:thresholds}
 \begin{tabular}{lccccc}
  \hline
  $\chi$ & $r(\chi)$ & Audibility & Listening & Relay & Mutual listening\\
  \hline
  $10^{-4}$ & 1.000 & 1.324(5) & 1.324(12) & 1.521(8) & 1.322(5) \\
  $10^{-3}$ & 0.999 & 1.318(5) & 1.347(18) & 1.516(24) & 1.331(5) \\
  $10^{-2}$ & 0.993 & 1.316(5) & 1.352(32) & 1.531(24) & 1.339(5) \\
  0.03 & 0.980 & 1.291(5) & 1.343(29) & 1.504(24) & 1.351(5) \\
  0.05 & 0.967 & 1.272(5) & 1.332(17) & 1.521(23) & 1.379(5) \\
  0.07 & 0.953 & 1.256(5) & 1.342(28) & 1.490(23) & 1.387(6) \\
  0.1 & 0.933 & 1.223(4) & 1.315(17) & 1.460(24) & 1.423(5) \\
  0.2 & 0.867 & 1.153(4) & 1.293(27) & 1.418(23) & 1.545(5) \\
  0.3 & 0.800 & 1.083(4) & 1.261(32) & 1.365(21) & 1.682(5) \\
  0.5 & 0.667 & 0.968(3) & 1.197(26) & 1.300(20) & 2.035(7) \\
  0.7 & 0.533 & 0.877(3) & 1.136(23) & 1.204(19) & 2.546(7) \\
  1 & 0.333 & 0.759(3) & 1.052(21) & 1.131(17) & 4.001(10) \\
  1.5 & 0.148 & 0.655(3) & 0.953(19) & 1.024(16) & 8.990(22) \\
  3 & 0.037 & 0.528(2) & 0.788(16) & 0.876(7) & 35.92(9) \\
  10 & 0.003 & 0.392(1) & 0.636(7) & 0.726(3) & 398.4(9) \\
  \hline
 \end{tabular}
\end{table*}

\begin{figure*}
\centering
\includegraphics[width=\textwidth]{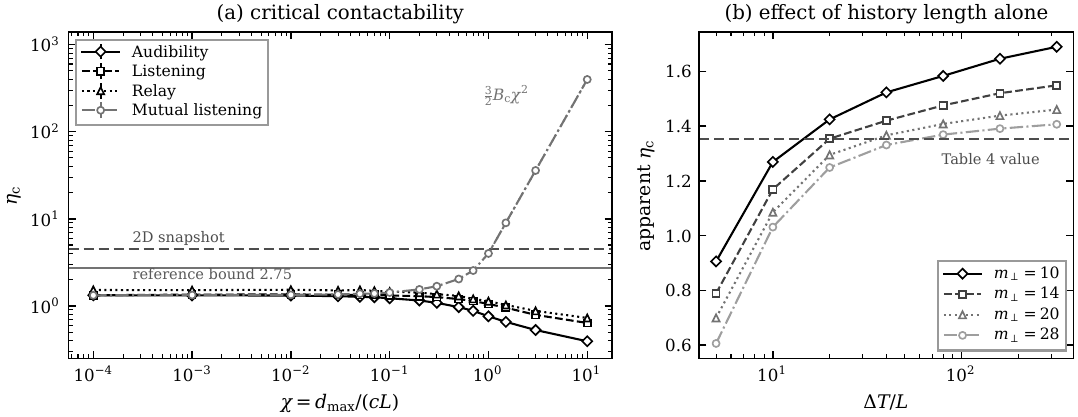}
\caption{
(a) Critical contactability $\eta_{\rm c}$ of the four networks
(Table~\ref{tab:thresholds}) as a function of the reach ratio $\chi$.
The dashed line marks the coeval threshold $\eta_{\rm c}=4.51$ of
Paper~II, the solid grey line the ideal-survey benchmark
$\eta(R_0)\le2.75$, and the dotted curve Equation~\eqref{eq:bicone2}.
(b) Apparent listening threshold at $\chi=0.01$ as a function of temporal
span $\Delta T$ for four spatial cross-sections $m_\perp$.
At fixed $m_\perp$ the apparent threshold rises throughout the range shown;
the level it reaches falls towards the tabulated threshold (dashed line) as
$m_\perp$ increases.
Uncertainties are those of Table~\ref{tab:thresholds}.
}
\label{fig:thresholds}
\end{figure*}

\subsection{Thresholds of the four networks}
\label{sec:threevalues}

Table~\ref{tab:thresholds} and Figure~\ref{fig:thresholds} summarize the
percolation thresholds of the four networks.
Four main conclusions emerge.

\emph{(i) The spacetime contact threshold is $\eta_{\rm c}\simeq1.3$, not
$4.5$.}
As $\chi\rightarrow0$, the listening, audibility, and mutual-listening
relations converge, yielding
\begin{equation}
\eta_{\rm c}(\chi\rightarrow0)=1.32\pm0.01.
\label{eq:etazero}
\end{equation}
This value is a factor of 3.4 below the coeval threshold
$\eta_{\rm c}=4.51$ adopted in Papers~I and II. Distributing civilizations
through time therefore lowers the contactability required for
Galaxy-spanning connectivity. The result reproduces the known undirected
$\chi\rightarrow0$ limit of \citet{Hyytia2012}; the new results are the dependence on $\chi$,
the distinction between contact relations, and the extension to a structured
Galactic disc.

\emph{(ii) All networks except mutual listening percolate more easily as
$\chi$ increases.}
Between $\chi=10^{-4}$ and $10$, the audibility threshold falls by about
70 per cent and the listening and relay thresholds by about half, whereas
the mutual-listening threshold increases by a factor of about $300$. For $\chi\gtrsim0.2$ the ordering 
$\eta_{\rm c}^{\rm aud} < \eta_{\rm c}^{\rm lis} < \eta_{\rm c}^{\rm relay} < \eta_{\rm c}^{\rm mut}$
remains unchanged.

\emph{(iii) Audibility and mutual listening become distinct connectivity
criteria.}
At $\chi=1$, audibility percolates at $\eta\simeq0.76$, whereas mutual
listening requires $\eta\simeq4.00$. A Galaxy-spanning audible population
therefore need not form a large mutual-listening network, and the separation
increases with $\chi$.

\emph{(iv) Relay connectivity overtakes mutual listening between
$\chi=0.1$ and $0.2$.}
For small reach ratios, mutual listening percolates more easily than relay
reachability. Above the crossover, relay connectivity becomes the less
demanding criterion. At $\chi=1$, relay percolation requires roughly
3.5 times lower contactability than mutual listening. The implications for
realistic SETI systems are discussed in Section~\ref{sec:discussion}.

\subsection{The mutual-listening branch at
\texorpdfstring{$\chi\ge1$}{chi >= 1}}
\label{sec:excluded}

The strong increase of the mutual-listening threshold in
Table~\ref{tab:thresholds} can be understood analytically for $\chi\ge1$.
When $\chi\ge1$, the detection range exceeds the light horizon,
$d_{\rm max}\ge cL$, whereas mutual listening always requires $d<cL$
(Equation~\ref{eq:mutual}). The additional detection range therefore
contributes audibility but not mutual listening. At fixed civilization
density and lifetime, $\langle k\rangle_{\rm mut} = \frac{2}{3}\pi n c^{2}L^{2}$,
which is independent of $d_{\rm max}$. The mutual-listening graph is
therefore the same at every $\chi\ge1$, so it percolates at a single
critical mean degree $B_{\rm c}^{\rm mut}$, a pure number. Using
Equation~\eqref{eq:mutviar} together with $r=1/(3\chi^{2})$ gives
\begin{equation}
\eta_{\rm c}^{\rm mut}
=
\frac{3}{2}
B_{\rm c}^{\rm mut}
\chi^{2},
\qquad
\chi\ge1.
\label{eq:bicone2}
\end{equation}
Thus the threshold grows exactly as $\chi^{2}$.
Measurements at $\chi=1$, 3 and 10 give
$B_{\rm c}^{\rm mut}=2.659\pm0.006$, so that
$\eta_{\rm c}^{\rm mut}=3.99\,\chi^{2}$, which reproduces the measured
thresholds to better than 1 per cent.

Building a more sensitive receiver
extends $d_{\rm max}$, but mutual listening still requires a separation
below $cL$, which the communicative lifetime alone fixes. At fixed lifetime,
therefore, the physical density of civilizations needed for a
Galaxy-spanning mutual-listening network does not change as the detection
range grows, even though the critical contactability rises as $\chi^{2}$.
The two statements agree because $\eta$ counts civilizations within
$d_{\rm max}$ while the links form only within the smaller radius $cL$.
Greater range does help one-way listening and relay connectivity. What it
cannot provide is reciprocity, and a completed reply is limited by the shorter
distance $cL/2$.

\subsection{The extra density required for relaying}
\label{sec:relaycost}

Section~\ref{sec:relaywindow} showed that relay reachability remains
substantial despite the finite residual lifetime.
A civilization forwards a message only during the part of its lifetime that
remains, so at a given contactability its relay out-degree is
$\eta(1+\ell/L)$ rather than the listening value $2\eta$, and a higher
density is needed before relay chains span the system.
Table~\ref{tab:thresholds} measures how much higher, by comparing the
listening and relay thresholds. The measured ratio is
\begin{equation}
\frac{\eta_{\rm c}^{\rm relay}}
     {\eta_{\rm c}^{\rm lis}}
\simeq
1.06\text{--}1.15,
\qquad
\text{mean }1.11,
\label{eq:relaycost}
\end{equation}
across five decades in $\chi$.

Relaying therefore requires a civilization density only about 10 per cent
higher than that implied by the corresponding listening network. At the
smallest $\chi$, where the listening threshold coincides with the undirected
one, the ratio reaches $1.521/1.324=1.15$, the same 15 per cent increase that
\citet{HyytiaOtt2013} report between the undirected and the causal model.
Both thresholds come from directed-network estimators that carry residual
finite-size and finite-time systematics (\ref{app:thresholds}). We therefore
describe the relay threshold as roughly 10 per cent higher than the
listening threshold, and no more precisely.

\subsection{How long a history is required}
\label{sec:span}

Papers~I and~II involved a single finite-size scale, the spatial extent of
the system. The spacetime formulation introduces a second scale: the number
of communicative generations represented by the simulated history,
$\Delta T/L$.
Increasing the temporal extent alone does not produce convergence. As shown
in Figure~\ref{fig:thresholds}(b), the apparent threshold keeps rising with
$\Delta T$ when the spatial cross-section is held fixed. A system of fixed
cross-section and unbounded duration is effectively one-dimensional. In this
Poisson model, such a system does not percolate at any finite density: an
infinitely long column contains, with probability one, temporal gaps longer
than the reach of any link. No path can cross these gaps
\citep{MeesterRoy1996}.

For the undirected networks, scaling the spatial and temporal extents together makes the sequence
converge, so that Equation~\eqref{eq:fss} applies. Directed
networks stay anisotropic and therefore need the survival-time criterion of
Section~\ref{sec:mcmethod}.

Two consequences follow. First, the required history length is determined by
the correlation scale of the network rather than by a simple message
crossing time. Second, short histories systematically underestimate the
threshold, making connectivity appear easier than it is in the asymptotic
limit. Although this bias decreases with increasing $\Delta T$, it remains
significant over long intervals.

The duration $\Delta T$ is therefore not only a numerical control
parameter. In a Galaxy of finite age it is also part of the physical
problem. The thresholds in Table~\ref{tab:thresholds} target the
long-history limit; for the directed networks, residual finite-size and
finite-time systematics remain as discussed in
\ref{app:thresholds}.


\section{The contact frontier in spacetime}
\label{sec:ghz}

Section~\ref{sec:thresholds} considered a uniform infinite medium. The
Galaxy is neither uniform nor infinite. Paper~II showed that spatial density
gradients replace a sharp contact boundary with a frontier of finite width.
This section extends that framework to the spacetime model.

\subsection{The structured disc and the Galactic habitable zone}
\label{sec:disc}

Paper~II generalized the uniform model by allowing civilization density to
vary with Galactocentric radius,
\begin{equation}
n(r)
=
\zeta(r)\,\Sigma_{\star}(r),
\qquad
\Sigma_{\star}(r)
=
\Sigma_{\star}(R_0)
e^{(R_0-r)/h_R},
\label{eq:ndisc}
\end{equation}
where $\Sigma_{\star}(r)$ is the stellar surface density, $h_R$ is the disc
scale length, and $\zeta(r)$ is the occurrence fraction of communicative
civilizations; a radially varying $\zeta(r)$ corresponds to a Galactic
habitable zone \citep[GHZ;][]{Gonzalez2001}.

Throughout this section we adopt the reference Galactic model of
Paper~II with $R_0=8.2$~kpc, $h_R=2.6$~kpc, a disc truncation radius
$R_{\rm tr}=20$~kpc, and $\eta(R_0)=1$. At the reference setting $\chi=1$
this gives $d_{\rm max}=cL=1.86$~kpc and $N=91$. For a uniform occurrence fraction,
\begin{equation}
\eta(r)
=
\pi n(r)d_{\rm max}^2
=
\eta(R_0)
e^{(R_0-r)/h_R},
\label{eq:etaexp}
\end{equation}
so the contactability field is a pure exponential.

Following Paper~II, we consider three representative GHZ models, in which
the civilization surface density $n(r)$ peaks at $2.2$~kpc
(Inner; \citealp{Prantzos2008,Gowanlock2011}), $7.2$~kpc
(Middle; \citealp{Lineweaver2004}), and $12.0$~kpc
(Outer; \citealp{Vukotic2016}).
Unless stated otherwise, however, the results below refer to the
uniform-$\zeta$ case, for which Equation~\eqref{eq:etaexp} applies.

\subsection{The contactability field when the two lengths differ}
\label{sec:fieldchi}

Paper~II defined the contactability field using the light horizon $cL$ as
the contact length. Treating the detection range separately gives
\begin{equation}
\eta(r)
=
\pi n(r)\,d_{\rm max}^{2}
=
\chi^{2}
\Bigl[\pi n(r)(cL)^{2}\Bigr],
\label{eq:fieldchi}
\end{equation}
where the bracketed term is the contactability field of Paper~II. The
factor $\chi^{2}$ changes only the normalization and leaves the radial shape
unchanged. Consequently, the methods developed in Paper~II apply directly
once the appropriate threshold $\eta_{\rm c}$ is specified.

For an exponential disc, changing the normalization shifts every frontier
radius by $2h_R\ln\chi$. The reach ratio therefore affects the frontier
through both this geometric shift and the threshold $\eta_{\rm c}$, while
differences between the four contact networks enter only through their
different threshold values.

\begin{table*}
 \centering
\caption{Frontier radii (kpc) for the reference disc model
[$\eta(R_0)=1$, $R_0=8.2$~kpc, $h_R=2.6$~kpc].
The first four columns give the
local-criterion radius from Equation~\eqref{eq:rfront} using the thresholds
of Table~\ref{tab:thresholds}; the $\chi\rightarrow0$ column uses the
$\chi=10^{-4}$ thresholds. A dash indicates a negative radius or, for the
coeval model, that the definition applies only at $\chi=1$.
The final two columns give the measured frontier at $\chi=1$
(Section~\ref{sec:frontmeasured}): $r_{50}$ is the radius where the
connected fraction reaches $g=0.5$, and
$w=r(g=0.1)-r(g=0.9)$ is the frontier width. Here $g$ denotes membership in
the largest component for the undirected networks and the coeval model, and
$g_{\rm seed}$ for the listening and relay networks.
All radii are computed at fixed $\eta(R_0)=1$ and therefore exclude the
geometric shift $2h_R\ln\chi$ discussed in
Section~\ref{sec:fieldchi}.}
 \label{tab:frontiers}
 \begin{tabular}{lcccccc}
  \hline
  Network & \multicolumn{4}{c}{Local criterion, Equation~(\ref{eq:rfront})} & \multicolumn{2}{c}{Measured at $\chi=1$}\\
  \cline{2-5}\cline{6-7}
          & $\chi \to 0$ & $\chi = 1$ & $\chi = 3$ & $\chi = 10$ & $r_{50}$ & $w$\\
  \hline
  Audibility                        & 7.47 & 8.92 & 9.86 & 10.63 & 10.22 & 3.77 \\
  Listening                         & 7.47 & 8.07 & 8.82 & 9.38  & 8.91  & 3.81 \\
  Relay                             & 7.11 & 7.88 & 8.54 & 9.03  & 8.70  & 3.64 \\
  Mutual listening                  & 7.47 & 4.59 & ---  & ---   & 5.52  & 3.37 \\
  coeval, two dimensions (Paper~II) & ---  & 4.28 & ---  & ---   & 6.49  & 3.94 \\
  \hline
 \end{tabular}
\end{table*}

\subsection{Frontier radii}
\label{sec:frontiers}

For an exponential contactability field, the frontier radius is
\begin{equation}
r(\eta_{\rm c})
=
R_{0}
+
h_{R}
\ln\!\left[
\frac{\eta(R_{0})}
     {\eta_{\rm c}}
\right],
\label{eq:rfront}
\end{equation}
as derived in Paper~II. Table~\ref{tab:frontiers} evaluates this expression
using the thresholds of Table~\ref{tab:thresholds} and the reference disc
model.

The principal effect of temporal structure is seen in the small-$\chi$
limit, where the coeval and spacetime models differ only through their
thresholds. Replacing the coeval value $\eta_{\rm c}=4.51$ with the
spacetime value $\eta_{\rm c}=1.32$ shifts the frontier outward by
\begin{equation}
\Delta r
=
h_{R}
\ln\!\left(\frac{4.51}{1.32}\right)
=
3.2~{\rm kpc},
\label{eq:dr}
\end{equation}
moving it from 4.28 to 7.47 kpc at fixed $\eta(R_0)$.
In the $\chi\rightarrow0$ limit the relay frontier lies $0.36$~kpc inside
the listening frontier, much smaller than the $3.2$~kpc shift from the
coeval to the spacetime framework.

At fixed $\eta(R_0)$, all frontiers except that of mutual listening move
outward as $\chi$ increases. The Sun lies inside a frontier when
$\eta(R_0)>\eta_{\rm c}$, which occurs above characteristic reach ratios of
approximately 0.43 for audibility, 1.2 for listening, and 1.7 for relay.
The mutual-listening frontier instead moves inward with increasing $\chi$.

Equation~\eqref{eq:rfront} provides only a local estimate of frontier
position. As shown in Paper~II, the transition occupies a broad radial band,
so the radii in Table~\ref{tab:frontiers} should be interpreted as
characteristic locations rather than sharp boundaries. Section~\ref{sec:frontmeasured}
compares these local estimates with measured frontier profiles.

\subsection{How wide the front is}
\label{sec:frontmeasured}

A spatial gradient produces a transition region of finite width rather than
a sharp percolation boundary. The Galactic density gradient is steep enough
that this region spans only about two detection ranges, so we measure its
width directly, as Paper~II did.

\begin{figure*}
\centering
\includegraphics[width=\textwidth]{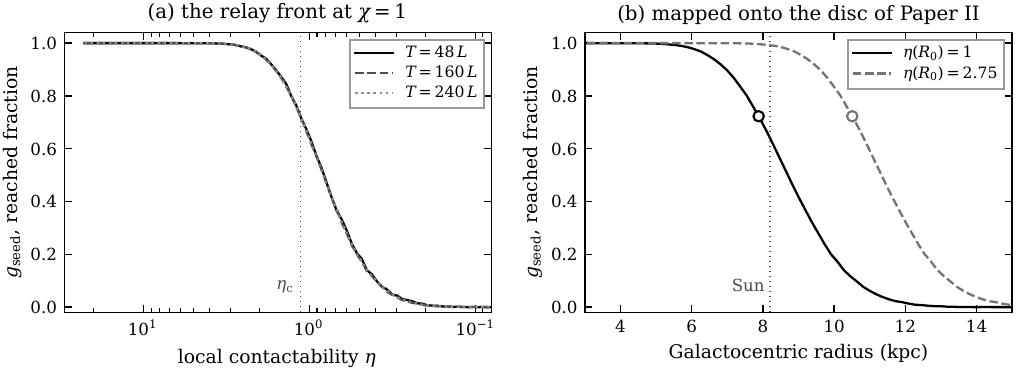}
\caption{
(a) Relay-connected fraction $g_{\rm seed}$ at $\chi=1$ for history lengths
$T=48$, $160$, and $240\,L$. The curves overlap. The dotted line marks the
relay threshold $\eta_{\rm c}=1.13$.
(b) The same profile as a function of Galactocentric radius for
$\eta(R_0)=1$ and $\eta(R_0)=2.75$. Open circles show the local criterion
(Equation~\ref{eq:rfront}); the dotted line marks the Sun.
}
\label{fig:front}
\end{figure*}

Using the exponential field of Equation~\eqref{eq:etaexp}, we measure the
relay-connected fraction $g_{\rm seed}(r)$ by propagating time-respecting
relay paths from the dense side of the disc. Every civilization within one
detection range of the dense edge is used as a seed, at any epoch, and
reachability is measured in the forward direction only. The profile is
accumulated over the interior of the simulated history, leaving out
civilizations within a few lifetimes of either temporal boundary. The
frontier width is defined throughout as $w=r(g=0.1)-r(g=0.9)$.
The resulting profile is shown in Figure~\ref{fig:front}. For the relay
network,
\begin{equation}
w_{\rm relay}
=
(1.96\pm0.01)\,d_{\rm max}
=
1.40\,h_R
=
3.64~{\rm kpc},
\label{eq:wmeas}
\end{equation}
measured at a history length $T=240\,L$. The width converges rapidly once
the temporal boundaries are excluded from the analysis.
This value is close to the coeval result of Paper~II
($3.89\pm0.03$~kpc). Distributing civilizations through time therefore
shifts the frontier position substantially while leaving its width nearly
unchanged.

The final two columns of Table~\ref{tab:frontiers} compare all networks at
$\chi=1$. Frontier positions vary by almost $5$~kpc, from
$5.5$~kpc for mutual listening to $10.2$~kpc for audibility. In contrast,
all measured widths lie between $3.37$ and $3.94$~kpc. The coeval width of
$3.94$~kpc in Table~\ref{tab:frontiers} is our own remeasurement; Paper~II
reported $3.89\pm0.03$~kpc for the same quantity. For a population with a
common lifetime and a common detection range, the width is therefore
controlled primarily by the Galactic density gradient rather than by the
contact relation itself in the reference disc model. We have not varied the
detection range or the civilization abundance widely enough to say how the
width behaves outside that model.

The measured frontiers lie systematically outside the local-criterion
radii of Table~\ref{tab:frontiers}, confirming the non-local effect already
identified in Paper~II. For the relay network at $\chi=1$, the local
criterion $\eta=\eta_{\rm c}$ corresponds to
$g_{\rm seed}\simeq0.72$, placing it well inside the transition region
rather than at its midpoint. The local criterion should therefore be
interpreted as a characteristic frontier location rather than as the
50\%-membership radius.

\subsection{Validity of the local criterion}
\label{sec:localcrit}

Neither correction discussed in Section~\ref{sec:recovery} significantly
affects the present results. The edge deficit is negligible because it lies
beyond all frontiers considered here. For the reference model, the exact
non-local contactability differs from the local estimate by only 4.4\% at
the Sun's Galactocentric radius, so the local criterion remains adequate. A substantial
correction appears only for the Outer GHZ model, where the non-local
treatment of Paper~II is required.

The thresholds of Section~\ref{sec:thresholds} are derived for an infinite
uniform medium and, for the directed networks, retain residual finite-size
and finite-time systematics. Additional finite-$N$ corrections may become
important for sufficiently small Galactic populations, but are not included
here.


\section{Discussion}
\label{sec:discussion}

Because $\chi$ is set by observational reach, the contact counts of
Section~\ref{sec:contactability} help to interpret SETI results. This section examines that question and then returns to the
assumptions of Section~\ref{sec:formulation}.

\subsection{The counts on the observational axis}
\label{sec:onaxis}

Evaluating the counts of Section~\ref{sec:contactability} at the reach
ratios of Table~\ref{tab:chiscale} gives three regimes. The listening count
is $2\eta$ throughout, and the reciprocity $r(\chi)$ of
Equation~\eqref{eq:recip} converts it into the others through
Equations~\eqref{eq:mutviar}, \eqref{eq:audviar} and~\eqref{eq:kreply}.

For $\chi\lesssim0.1$, corresponding to leakage-like technosignatures,
reciprocity stays close to unity and the listening, audibility and
mutual-listening relations are nearly identical. All three counts approach
$2\eta$, so the principal effect of relaxing coevality is the universal
factor-of-two increase relative to Papers~I and II.

At intermediate reach ratios the relations remain comparable but
increasingly distinct. At the Arecibo-class radar range ($\chi=0.192$),
$r=0.87$, so mutual listening is $1.7\eta$ against the listening count
$2\eta$. At the SKA2 design goal ($\chi=0.607$), $r=0.60$: mutual listening
has fallen to $1.2\eta$ and the reply-closing count to $0.91\eta$, while the
listening count is unchanged. Reciprocity passes one half at $\chi=0.75$.

Above $\chi=1$ the relations diverge rapidly. At the planetary-radar reach
ratio $\chi=1.97$, $r=0.086$: the mutual-listening count is $0.17\eta$ and
the reply-closing count $0.086\eta$, while the listening count remains
$2\eta$, about twenty-three times the latter. At that reach, being detectable and being able to hold a conversation are
different conditions.

\subsection{What a non-detection constrains}
\label{sec:nondetection}

Throughout this section we compare the network thresholds with the reference
bound of Paper~II, $\eta(R_0)\le2.75$. That bound follows from requiring that
our own non-detection have a probability of at least 5 per cent, and it was
derived at $d_{\rm max}=cL$. Because it also assumes a survey that detects
every transmitter within the detection range, it is a benchmark rather than a
limit set by any actual search. The weaker Poisson limit
$\eta\le\ln(1/0.05)=3.0$, which follows from the same 5 per cent requirement
without the statistical model, gives a similar benchmark.

Two observationally relevant results follow immediately. First, at the
reference bound, the lifetime listening count is $\langle k\rangle_{\rm lis} = 2\eta = 5.5$,
independent of the reach ratio $\chi$. This is an average over civilization
appearances, not a prediction for a particular observer: half of the count
consists of signals that are already arriving when a lifetime begins.
Second, the number of contacts that
complete a two-way exchange is much smaller. Requiring that those
civilizations hear us in return leaves $1.8$ at $\chi=1$ and $0.5$ at the
planetary-radar value of Table~\ref{tab:chiscale}; requiring that a reply
also arrive leaves $0.9$ and $0.2$. Contact may therefore be common even
when completed conversations are rare.

A non-detection primarily constrains the single-epoch count
$\langle k\rangle_{\rm obs}=\eta$, not the lifetime-integrated listening
count. The duration of existing SETI observations is short compared with the
reference communicative lifetime $L$, so the difference between a
single-epoch and finite-duration observer is negligible. Both counts already
include transmitters that stopped long ago, because a signal outlives its
source in either case; the factor of two between them comes from how long the
receiver lasts, not from counting civilizations that are already extinct
(Section~\ref{sec:compare}). Consequently, a null result constrains the same
quantity $\eta$ as in Papers~I and II.

The principal consequence of the spacetime framework is that the
connectivity threshold can now fall below the observational benchmark. In
the coeval framework this was impossible because
$\eta_{\rm c}=4.51>2.75$. In the spacetime framework, however,
\begin{equation}
\eta_{\rm c}(\chi)
\le
\eta(R_0)
\le
2.75 ,
\label{eq:window}
\end{equation}
defines a non-empty interval for listening, audibility, and relay across
the full range of $\chi$ considered here. Mutual listening is the exception:
its threshold exceeds 2.75 for $\chi\gtrsim0.75$, closing the window. The
conclusion does not depend on the exact value of the bound, because the
thresholds of the other three networks lie below both 2.75 and the Poisson
value 3.0 throughout Table~\ref{tab:thresholds}.

Thus, under the same reference benchmark, the local connectivity criterion
and the non-detection constraint can be satisfied simultaneously in the
spacetime framework, whereas they cannot in the coeval framework. This
establishes that the two requirements are compatible. It does not establish
that the Sun belongs to a Galaxy-spanning connected component: membership is
a non-local property, set by the surroundings out to the correlation scale
of the network rather than by the local contactability alone
(Section~\ref{sec:localcrit}).

For a fixed civilization density and lifetime, greater sensitivity increases
the number of civilizations we could hear. Once the detection range exceeds
$cL/2$, however, it no longer increases the mean number of reply-closing
contacts, which are limited by the time available for a return signal.

These conclusions depend on the assumptions used to derive the benchmark. A bound
derived from the actual sky, frequency, and time coverage of SETI surveys
would provide a more realistic observational test.

\subsection{What a detection would imply}
\label{sec:detection}

For a signal detected at a random observation epoch, the probability that
the transmitting civilization remains active is
\begin{equation}
P(\mathrm{still\ transmitting}\mid\mathrm{detected})
=
r(\chi),
\label{eq:survive}
\end{equation}
where $r(\chi)$ is the reciprocity of
Equation~\eqref{eq:recip}. This result assumes both A2 and A5, and the
non-uniform Galactic field of Section~\ref{sec:ghz} adds a distance-dependent
weight that $\chi$ alone does not describe.
Reciprocity therefore has a direct observational meaning. For
$\chi\ll1$, a detected civilization is almost certainly contemporary. For
$\chi\gtrsim1$, most detected signals originate from civilizations that are
already inactive. At $\chi=1.97$, the probability of continued
transmission is only $0.086$.

A detection therefore establishes that a signal was emitted, but not that a
two-way exchange remains possible. Reciprocity and reply-closing become
increasingly restrictive as $\chi$ increases.

\subsection{The five assumptions}
\label{sec:assumptions}

The spacetime model of Section~\ref{sec:formulation} depends on five
assumptions. Their effects on contact counts and percolation thresholds are
not identical; in general, thresholds are the more sensitive quantities.

\paragraph*{A1. Common transmitting and listening interval}
If the listening duration is $b$ instead of $L$,
\begin{equation}
\langle k\rangle_{\rm lis}
=
\eta\left(1+\frac{b}{L}\right),
\label{eq:listenb}
\end{equation}
recovering $\langle k\rangle_{\rm obs}=\eta$ for $b=0$ and the present
model for $b=L$. A finite listening interval is also what gives rise to
reciprocity, mutual listening, and finite temporal connection regions.

\paragraph*{A2. A common communicative lifetime}
As shown in Sections~\ref{sec:listening} and~\ref{sec:audmut}, the
listening count $\langle k\rangle_{\rm lis}=2\eta$ and the sum rule of
Equation~\eqref{eq:sumrule} remain valid for arbitrary lifetime
distributions, provided lifetime and $d_{\rm max}^{2}$ are uncorrelated. Relaxing A2 therefore redistributes contacts among the
different relations without changing their total number. Thresholds are more
sensitive because they depend on the detailed shape of the connection
regions. Paper~IV explores this case together with A5.

\paragraph*{A3. Stationary appearances}
Contact counts depend only on the finite temporal extent of the link
conditions and therefore require stationarity only over intervals of order
$(1+\chi)L$. Connectivity depends on the full appearance history and can be
modified substantially by long-term variations in civilization formation
rate.

\paragraph*{A4. Universal retransmission}
Assumption A4 states that civilizations can relay messages whenever they
receive them. Relay counts should therefore be interpreted as upper bounds.
If only a fraction $p$ of recipients retransmit, relay branching factors
scale approximately as $p$ and the corresponding thresholds as $1/p$. This
assumption is likely to have the largest quantitative effect on relay
connectivity.

\paragraph*{A5. A common detection range and isotropic emission}
A distribution of detection ranges replaces the single parameter $\chi$ by a
population distribution while leaving $\eta$ meaningful as a population
average. Directional transmission affects the contact relations differently:
for a beam covering a sky fraction $\phi$, listening scales as $\phi$,
whereas mutual listening scales as $\phi^2$. Beaming therefore suppresses
reciprocity and widens the difference between audibility and mutual
listening.

Taken together, these assumptions determine the numerical values of the
contact counts and thresholds. The qualitative distinction between
listening, audibility, mutual listening, and relay communication does not
depend on them. Once civilizations have finite lifetimes and are distributed
through spacetime, these contact relations become intrinsically different.
The quantitative separation between their thresholds depends on the
population model.

\subsection{What the Fermi paradox assumes}
\label{sec:paradox}

The Fermi paradox is often framed as a comparison between the expected
number of Galactic civilizations and the absence of any detection. In that
form, one quantity, the abundance $N$, is taken to decide whether the
silence is surprising. The
present series separates this connection into several distinct quantities:
the abundance of civilizations $N$, the local contactability $\eta$, the
existence of a Galaxy-spanning communication network, and detectability by a
particular observer at a particular epoch.

These quantities are not interchangeable. A large abundance need not give a
large contactability, and typical contact begins well below the threshold
for a Galaxy-spanning network (Paper~I); local contactability does not
determine membership in that network (Paper~II); and the present paper shows
that distributing civilizations through time lowers the threshold below the
reference non-detection benchmark while leaving the benchmark itself
unchanged.

Detectability introduces a further layer. Whether a civilization is
detectable from Earth depends not only on the properties of the Galactic
population but also on transmitter power, observing sensitivity, survey
coverage, and transmission behaviour. Transmitter power and receiver
sensitivity enter the model through $d_{\rm max}$; survey coverage and the
detailed transmission behaviour are not modelled explicitly.

This series therefore does not resolve the Fermi paradox. It shows that
abundance alone does not determine contactability, network connectivity, or
detectability. The silence of the Galaxy, while still unexplained,
therefore does not by itself tell us whether the Galaxy contains a connected
communication network.


\section{Conclusions}
\label{sec:conclusions}

Papers~I and~II treated interstellar contact with a single length, the light
horizon $cL$, and assumed that civilizations in contact were contemporaries.
This paper separates the detection range $d_{\rm max}$ from that length and
treats civilizations as finite-lifetime events in spacetime. Contact then
splits into four relations, with a fifth condition deciding whether a reply
returns. We derive their contact counts analytically and estimate their
connectivity thresholds by simulation.

The principal result is that distributing civilizations through time lowers
the connectivity threshold below the reference non-detection benchmark of
Paper~II. In the coeval framework the local connectivity criterion
($\eta_{\rm c}=4.51$) exceeds the benchmark ($\eta(R_{0})\le2.75$), so the
two cannot be satisfied together. In spacetime the small-$\chi$ listening
threshold falls to $\eta_{\rm c}=1.32\pm0.01$, a factor of $3.4$, and both
can. This establishes consistency, not membership: whether the Sun belongs
to a Galactic connected component remains a non-local question.

Four results support this. Distributing finite lifetimes through time
doubles the contact count: the mean lifetime listening count is
$\langle k\rangle_{\rm lis}=2\eta$ at every $\chi$, and remains so for any
lifetime distribution at a common detection range. Audibility and mutual
listening then sum to a fixed total of $4\eta$; for the common-lifetime
population studied here, their division is set by the reciprocity $r(\chi)$.
A detected civilization is still transmitting with probability $r(\chi)$,
which is $0.086$ at the planetary-radar reach ratio. The relay threshold is
only $6$ to $15$ per cent above the listening threshold. In the structured Galactic
disc the local-criterion frontier moves outward by $3.2$~kpc, while the
relay front stays about as broad as the coeval one, $3.64$ against
$3.94$~kpc.

For SETI, the central implication is that detectability, audibility and
reciprocal communication are distinct conditions. A null detection still
constrains $\eta$, but under the reference benchmark a civilization can hear
up to $5.5$ others on average over a communicative lifetime, while the mean
reply-closing count falls below one once $\chi\gtrsim1$: contact may be
common even when completed exchanges are rare. More broadly, abundance,
contactability, connectivity and detectability are separate conditions, and
Papers~I--III measure how far apart the first three lie
(Section~\ref{sec:paradox}).

Emission is assumed isotropic here and every recipient is assumed to relay;
of the assumptions made, the relay fraction is the one most likely to move
the numerical thresholds. Paper~IV relaxes the common lifetime and the
common detection range.

\appendix
\setcounter{table}{0}
\setcounter{figure}{0}

\section{Numerical methods and validation}
\label{app:thresholds}

This appendix sets out how the listening and relay columns of
Table~\ref{tab:thresholds} were obtained and how their reliability was
checked. It describes a measurement, not a study of the critical behaviour:
no universality class is assumed or determined.

\subsection*{Why a different estimator is needed}

The finite-size extrapolation of Equation~\eqref{eq:fss} assumes that the
spatial and temporal extents of the box can be scaled together. For the
directed networks this does not hold over the sizes reached here. At fixed
aspect ratio the apparent relay threshold is not monotonic in box size, and
at fixed spatial extent it keeps rising as the simulated history is
lengthened. A different estimator is therefore required.

\subsection*{The survival-time criterion}

Civilizations are generated as in Section~\ref{sec:mcmethod}, with periodic
spatial boundaries and open temporal boundaries. Every civilization
appearing in the first generation acts as a source, reachability is
propagated by repeated application of the listening or relay rules, and the
longest time reached is recorded for each realization. The median of that
quantity over realizations is $t_{50}$. Near the transition we use the
approximate scaling criterion
\begin{equation}
t_{50}
\propto
m^{z}
\left(2\ln m + C\right),
\label{eq:t50}
\end{equation}
where $z$ is the dynamic exponent and $C$ a constant of order unity, the
logarithm coming from the $O(m^{2})$ sources in the first generation. The logarithmic slope between two box sizes is
therefore
\begin{equation}
Z
\equiv
\frac{\ln[t_{50}(m_2)/t_{50}(m_1)]}
     {\ln(m_2/m_1)}
\;\simeq\;
z+\frac{2}{2\ln\sqrt{m_1m_2}+C} ,
\label{eq:zcrit}
\end{equation}
and the threshold is the density at which $Z$ reaches that value. We use
$C=\ln2$ throughout, which puts the offset at $0.29$ above $z$ for
$m_1=16$ and $m_2=32$. The offset matters, since reading the criterion as
$Z=z$ would place the threshold elsewhere; it is checked against a known
threshold below.

Production runs used $m_1=16$ and $m_2=32$, an initial horizon
$T=\max(500,80\chi)L$, and 400 to 700 realizations at each of six densities
spanning about $\pm8$ per cent around each transition. Relay thresholds are
evaluated at $z=1.85$, the midpoint of the bracket $1.70\le z\le2.00$, and
listening thresholds are the mean of the crossings at $z=1$ and $z=2$. The
bracket is a range we adopt rather than a measured exponent, and the
tabulated thresholds change by less than their quoted uncertainties across
it. The quoted uncertainty is in each case the larger of the density-bracket
half-width and the spread over the adopted range of $z$. It is numerical,
not a statistical $1\sigma$ interval.

A realization whose reachable set is still alive at the end of the box is
given the value $T$, and the median is treated as censored when this happens
in more than half the realizations. A censored median does not by itself
show that the density is supercritical. It shows only that
$t_{50}(m_{2})\ge T$, which gives the lower bound
\begin{equation}
Z \;\ge\; \frac{\ln\left[T/t_{50}(m_{1})\right]}
                {\ln\left(m_{2}/m_{1}\right)} ,
\label{eq:zbound}
\end{equation}
useful only when $t_{50}(m_{1})$ is itself uncensored and well below the
horizon. A censored point is therefore accepted as the upper end of a
bracket only if Equation~\eqref{eq:zbound} already exceeds the target value
of $Z$, or if a repeat run with a longer horizon gives a measured $Z$ above
it.

Three cases at large reach ratio needed longer horizons. Relay at
$\chi=3$ was remeasured at $T=1.6\times10^{4}L$, with the crossing
interpolated at $z=1.85$ inside the density bracket $[0.871,0.884]$. Relay
at $\chi=10$ was remeasured at $T=3.2\times10^{4}L$, where the upper end of
the bracket $[0.723,0.728]$ was a censored point accepted under
Equation~\eqref{eq:zbound}, so the midpoint was taken. Listening at
$\chi=10$ was remeasured at the same horizon and interpolated inside
$[0.632,0.645]$, averaging the crossings at $z=1$ and $z=2$. These brackets
are the sampled density intervals containing the crossing, not the range of
the resulting estimate.

\subsection*{Validation}

The estimator was checked against the three benchmark problems named in
Section~\ref{sec:mcmethod}. For the three-dimensional unit-ball problem we
obtain a critical mean degree of $2.730\pm0.002$ against the reference value
$2.7351$ \citep{LorenzZiff2001}, where the quoted interval is the
block-to-block scatter of the estimator. The difference of $0.19$ per cent
is a small residual offset beyond that scatter. The other two benchmarks are
reproduced within the quoted uncertainties: the undirected
$\chi\rightarrow0$ limit of \citet{Hyytia2012}, $\eta_{\rm c}=1.3248(4)$ in
the present normalization, is returned as $1.324$, and the causal model of
\citet{HyytiaOtt2013}, $\eta_{\rm c}=1.515$, as $1.521\pm0.008$. The
undirected $\chi\rightarrow0$ benchmark also checks the offset in
Equation~\eqref{eq:zcrit}: at its known threshold the measured $Z$ is
$1.29$, matching the target $1+0.29$, so the correction has the stated size.

A further check compared two independent pairs of box sizes at $\chi=1$,
where the finite propagation time matters. The relay threshold is
$1.131\pm0.017$ from $(m_1,m_2)=(16,32)$ and $1.108\pm0.016$ from
$(24,48)$. The two agree within their uncertainties, and the difference of
2 per cent indicates a residual sensitivity to box size at $\chi=1$. The
intervals quoted in Table~\ref{tab:thresholds} describe the resolution of
the density grid and the dependence on the adopted range of $z$; they are
not a measurement of that size sensitivity, which we have not quantified at
other reach ratios. The listening and relay entries are therefore numerical
estimates for an infinite uniform medium, not exact critical densities.

\section*{Declaration of generative AI and AI-assisted technologies in the writing process}
The author used Claude (Anthropic), a large language model, to assist with the development of the manuscript, 
including the drafting and revision of text, improvement of English expression, refinement of argument structure, and preparation of figure concepts. 
The author also used Microsoft 365 Copilot (Microsoft Corporation), a large language model, to obtain a critical review of the manuscript, to perform independent checks of the results, and to assist with the drafting of the corresponding revised passages.
The author further used ChatGPT (OpenAI), a large language model, to assist with the implementation of the numerical codes used in this work, with the analysis of their output, and with revision of the English text.
All content generated with AI assistance was critically reviewed, edited, and verified by the author. 
The author assumes full responsibility for the content of the manuscript, including its analyses, interpretations, and conclusions.

\section*{Declaration of competing interest}
The author declares that he has no known competing financial interests or personal relationships
that could have appeared to influence the work reported in this paper.

\section*{Data availability}
The code and data supporting this study are available from the author upon reasonable request.



\bibliographystyle{elsarticle-num-names} 
\bibliography{GCNT3}

@misc{GCNT1,
  author        = {Tsumura, Kohji},
  title         = "{Contactability and Connectivity in Galactic Civilization Networks: A Network-Theoretic Framework for SETI}",
  howpublished  = {Zenodo},
  month         = oct,
  year          = 2026,
  doi           = {10.5281/zenodo.23190278},
}

@article{GCNT2,
    author = {Tsumura, Kohji},
    title = {The contact frontier in Galactic civilization networks: radial structure and the Galactic habitable zone},
    journal = {Monthly Notices of the Royal Astronomical Society},
    pages = {stag2021},
    year = {2026},
    month = {10},
    issn = {0035-8711},
    doi = {10.1093/mnras/stag2021},
    url = {https://doi.org/10.1093/mnras/stag2021}
}

@misc{GCNT4,
  author        = {Tsumura, Kohji},
  title         = "{Lifetime Distributions in Galactic Civilization Networks: A Hierarchy of Moments and Connectivity Thresholds}",
  howpublished  = {Zenodo},
  month         = oct,
  year          = 2026,
  doi           = {10.5281/zenodo.23191580},
}

@Article{Tsumura2020,
author={Tsumura, Kohji},
title="{Estimating survival probability using the terrestrial extinction history for the search for extraterrestrial life}",
journal={Scientific Reports},
year={2020},
month={Jul},
day={30},
volume={10},
number={1},
pages={12795},
issn={2045-2322},
doi={10.1038/s41598-020-69724-2},
url={https://doi.org/10.1038/s41598-020-69724-2}
}

@INCOLLECTION{Drake1965,
       author = {{Drake}, Frank D.},
        title = "{The Radio Search for Intelligent Extraterrestrial Life}",
    booktitle = {Current aspects of exobiology},
         year = 1965,
       editor = {{Mamikunian}, Gregg and {Briggs}, Michael H.},
        pages = {323-345},
       adsurl = {https://ui.adsabs.harvard.edu/abs/1965cae..book..323D}
}

@article{DallChristensen2002,
  title = "{Random geometric graphs}",
  author = {Dall, Jesper and Christensen, Michael},
  journal = {Phys. Rev. E},
  volume = {66},
  issue = {1},
  pages = {016121},
  numpages = {9},
  year = {2002},
  month = {Jul},
  publisher = {American Physical Society},
  doi = {10.1103/PhysRevE.66.016121},
  url = {https://link.aps.org/doi/10.1103/PhysRevE.66.016121}
}

@article{GrimaldiMarcy2018,
author = {Claudio Grimaldi  and Geoffrey W. Marcy },
title = "{Bayesian approach to SETI}",
journal = {Proceedings of the National Academy of Sciences},
volume = {115},
number = {42},
pages = {E9755-E9764},
year = {2018},
doi = {10.1073/pnas.1808578115},
URL = {https://www.pnas.org/doi/abs/10.1073/pnas.1808578115}}

@article{Kreifeldt1971,
title = "{A formulation for the number of communicative civilizations in the galaxy}",
journal = {Icarus},
volume = {14},
number = {3},
pages = {419-430},
year = {1971},
issn = {0019-1035},
doi = {10.1016/0019-1035(71)90011-X},
url = {https://www.sciencedirect.com/science/article/pii/001910357190011X},
author = {J.G. Kreifeldt}
}

@ARTICLE{Wallenhorst1981,
       author = {{Wallenhorst}, S.~G.},
        title = "{The Drake Equation Reexamined}",
      journal = {Quarterly Journal of the Royal Astronomical Society},
         year = 1981,
        month = dec,
       volume = {22},
        pages = {380},
       adsurl = {https://ui.adsabs.harvard.edu/abs/1981QJRAS..22..380W}
}

@article{Cirkovic2004,
author = {Milan M. \'{C}irkovi\'{c}},
title ="{The Temporal Aspect of the Drake Equation and SETI}",
journal = {Astrobiology},
volume = {4},
number = {2},
pages = {225-231},
year = {2004},
doi = {10.1089/153110704323175160},
URL = { https://journals.sagepub.com/doi/abs/10.1089/153110704323175160},
}

@article{Lineweaver2001,
title = "{An Estimate of the Age Distribution of Terrestrial Planets in the Universe: Quantifying Metallicity as a Selection Effect}",
journal = {Icarus},
volume = {151},
number = {2},
pages = {307-313},
year = {2001},
issn = {0019-1035},
doi = {10.1006/icar.2001.6607},
url = {https://www.sciencedirect.com/science/article/pii/S0019103501966072},
author = {Charles H. Lineweaver}
}

@article{Rushby2013,
author = {Andrew J. Rushby and Mark W. Claire and Hugh Osborn and Andrew J. Watson},
title ="{Habitable Zone Lifetimes of Exoplanets around Main Sequence Stars}",
journal = {Astrobiology},
volume = {13},
number = {9},
pages = {833-849},
year = {2013},
doi = {10.1089/ast.2012.0938},
URL = { https://journals.sagepub.com/doi/abs/10.1089/ast.2012.0938}
}

@article{WolfToon2015,
author = {Wolf, E. T. and Toon, O. B.},
title = "{The evolution of habitable climates under the brightening Sun}",
journal = {Journal of Geophysical Research: Atmospheres},
volume = {120},
number = {12},
pages = {5775-5794},
doi = {10.1002/2015JD023302},
url = {https://agupubs.onlinelibrary.wiley.com/doi/abs/10.1002/2015JD023302},
year = {2015}
}

@article{FreemanLampton1975,
title = "{Interstellar archaeology and the prevalence of intelligence}",
journal = {Icarus},
volume = {25},
number = {2},
pages = {368-369},
year = {1975},
issn = {0019-1035},
doi = {10.1016/0019-1035(75)90032-9},
url = {https://www.sciencedirect.com/science/article/pii/0019103575900329},
author = {J. Freeman and M. Lampton}
}

@article{Grimaldi2018,
doi = {10.1088/1538-3873/aaac6a},
url = {https://doi.org/10.1088/1538-3873/aaac6a},
year = {2018},
month = {mar},
publisher = {The Astronomical Society of the Pacific},
volume = {130},
number = {987},
pages = {054101},
author = {Grimaldi, Claudio and Marcy, Geoffrey W. and Tellis, Nathaniel K. and Drake, Frank},
title = "{Area Coverage of Expanding E.T. Signals in the Galaxy: SETI and Drake's N}",
journal = {Publications of the Astronomical Society of the Pacific},
}

@article{Balbi2018,
author = {Amedeo Balbi},
title ="{The Impact of the Temporal Distribution of Communicating Civilizations on Their Detectability}",
journal = {Astrobiology},
volume = {18},
number = {1},
pages = {54-58},
year = {2018},
doi = {10.1089/ast.2017.1652},
URL = { https://journals.sagepub.com/doi/abs/10.1089/ast.2017.1652},
}

@article{Wandel2015, 
title="{On the abundance of extraterrestrial life after the Kepler mission}", 
volume={14}, 
DOI={10.1017/S1473550414000767}, 
number={3}, 
journal={International Journal of Astrobiology}, 
author={Wandel, Amri}, 
year={2015}, 
pages={511-516}
}

@Article{CirkovicVukotic2008,
author={{\'{C}}irkovi{\'{c}}, Milan M.
and Vukoti{\'{c}}, Branislav},
title="{Astrobiological Phase Transition: Towards Resolution of Fermi's Paradox}",
journal={Origins of Life and Evolution of Biospheres},
year={2008},
month={Dec},
day={01},
volume={38},
number={6},
pages={535-547},
issn={1573-0875},
doi={10.1007/s11084-008-9149-y},
url={https://doi.org/10.1007/s11084-008-9149-y}
}

@article{Maccone2010,
title = "{The Statistical Drake Equation}",
journal = {Acta Astronautica},
volume = {67},
number = {11},
pages = {1366-1383},
year = {2010},
note = {{Special Issue on Searching for Life Signatures}},
issn = {0094-5765},
doi = {10.1016/j.actaastro.2010.05.003},
url = {https://www.sciencedirect.com/science/article/pii/S0094576510001499},
author = {Claudio Maccone}
}

@article{Smith2009, 
title="{Broadcasting but not receiving: density dependence considerations for SETI signals}", 
volume={8}, 
DOI={10.1017/S1473550409990097}, 
number={2}, 
journal={International Journal of Astrobiology}, 
author={Smith, Reginald D.}, 
year={2009}, 
pages={101-105}
}

@article{ForganRice2010, 
title="{Numerical testing of the Rare Earth Hypothesis using Monte Carlo realization techniques}", 
volume={9}, 
DOI={10.1017/S1473550410000030}, 
number={2}, 
journal={International Journal of Astrobiology}, 
author={Forgan, Duncan H. and Rice, Ken}, 
year={2010}, 
pages={73-80}
}

@Article{Grimaldi2017,
author={Grimaldi, Claudio},
title="{Signal coverage approach to the detection probability of hypothetical extraterrestrial emitters in the Milky Way}",
journal={Scientific Reports},
year={2017},
month={Apr},
day={12},
volume={7},
number={1},
pages={46273},
issn={2045-2322},
doi={10.1038/srep46273},
url={https://doi.org/10.1038/srep46273}
}

@article{Siemion2015,
  author = "Siemion, Andrew  and  Benford, James  and  Cheng-Jin, Jin  and  Chennamangalam, Jayanth  and  Cordes, James M.  and  Falcke, Heino  and  Garrington, Simon T.  and  Garrett, Michael A.  and  Gurvits, Leonid  and  Hoare, Melvin  and  Korpela, Eric  and  Lazio, Joseph  and  Messerschmitt, David  and  Morrison, Ian  and  O'Brien, Tim  and  Paragi, Zsolt  and  Penny, Alan  and  Spitler, Laura  and  Tarter, Jill  and  Werthimer, Dan",
  title = "{Searching for Extraterrestrial Intelligence with the Square Kilometre Array}",
  doi = "10.22323/1.215.0116",
  journal = "Proceedings of Science",
  year = 2015,
  volume = "AASKA14",
  pages = "116"
}

@article{Sheikh2025,
doi = {10.3847/1538-3881/ada3c7},
url = {https://doi.org/10.3847/1538-3881/ada3c7},
year = {2025},
month = {feb},
publisher = {The American Astronomical Society},
volume = {169},
number = {2},
pages = {118},
author = {Sheikh, Sofia Z. and Huston, Macy J. and Fan, Pinchen and Wright, Jason T. and Beatty, Thomas and Martini, Connor and Kopparapu, Ravi and Frank, Adam},
title = "{Earth Detecting Earth: At What Distance Could Earth's Constellation of Technosignatures Be Detected with Present-day Technology?}",
journal = {The Astronomical Journal}
}

@article{Smith2023, 
title="{Communicating extraterrestrial intelligence (CETI) interaction models based on the Drake equation}", 
volume={22}, 
DOI={10.1017/S1473550422000404}, 
number={2}, 
journal={International Journal of Astrobiology}, 
author={Smith, Reginald D.}, 
year={2023}, 
pages={157-165}
}

@Article{Dijkstra1959,
author={Dijkstra, E. W.},
title="{A note on two problems in connexion with graphs}",
journal={Numerische Mathematik},
year={1959},
month={Dec},
day={01},
volume={1},
number={1},
pages={269-271},
issn={0945-3245},
doi={10.1007/BF01386390},
url={https://doi.org/10.1007/BF01386390}
}

@article{HolmeSaramaki2012,
title = "{Temporal networks}",
journal = {Physics Reports},
volume = {519},
number = {3},
pages = {97-125},
year = {2012},
note = {{Temporal Networks}},
issn = {0370-1573},
doi = {10.1016/j.physrep.2012.03.001},
url = {https://www.sciencedirect.com/science/article/pii/S0370157312000841},
author = {Petter Holme and Jari Saram\"{a}ki}
}

@article{GarlaschelliLoffredo2004,
  title = "{Patterns of Link Reciprocity in Directed Networks}",
  author = {Garlaschelli, Diego and Loffredo, Maria I.},
  journal = {Physical Review Letters},
  volume = {93},
  issue = {26},
  pages = {268701},
  numpages = {4},
  year = {2004},
  month = {Dec},
  publisher = {American Physical Society},
  doi = {10.1103/PhysRevLett.93.268701},
  url = {https://link.aps.org/doi/10.1103/PhysRevLett.93.268701}
}

@article{LorenzZiff2001,
    author = {Lorenz, Christian D. and Ziff, Robert M.},
    title = "{Precise determination of the critical percolation threshold for the three-dimensional “Swiss cheese” model using a growth algorithm}",
    journal = {The Journal of Chemical Physics},
    volume = {114},
    number = {8},
    pages = {3659-3661},
    year = {2001},
    month = {02},
    issn = {0021-9606},
    doi = {10.1063/1.1338506},
    url = {https://doi.org/10.1063/1.1338506},
}

@article{NewmanZiff2000,
  title = "{Efficient Monte Carlo Algorithm and High-Precision Results for Percolation}",
  author = {Newman, M. E. J. and Ziff, R. M.},
  journal = {Physical Review Letters},
  volume = {85},
  issue = {19},
  pages = {4104--4107},
  numpages = {0},
  year = {2000},
  month = {Nov},
  publisher = {American Physical Society},
  doi = {10.1103/PhysRevLett.85.4104},
  url = {https://link.aps.org/doi/10.1103/PhysRevLett.85.4104}
}

@article{Wang2013,
  title = "{Bond and site percolation in three dimensions}",
  author = {Wang, Junfeng and Zhou, Zongzheng and Zhang, Wei and Garoni, Timothy M. and Deng, Youjin},
  journal = {Physical Review E},
  volume = {87},
  issue = {5},
  pages = {052107},
  numpages = {8},
  year = {2013},
  month = {May},
  publisher = {American Physical Society},
  doi = {10.1103/PhysRevE.87.052107},
  url = {https://link.aps.org/doi/10.1103/PhysRevE.87.052107}
}

@incollection{Prantzos2008,
author="Prantzos, Nikos",
editor="Botta, Oliver
and Bada, Jeffrey L.
and Gomez-Elvira, Javier
and Javaux, Emmanuelle
and Selsis, Franck
and Summons, Roger",
title="{On the ``Galactic Habitable Zone''}",
bookTitle="Strategies of Life Detection",
year="2008",
publisher="Springer US",
address="Boston, MA",
pages="313--322",
isbn="978-0-387-77516-6",
doi="10.1007/978-0-387-77516-6_22",
url="https://doi.org/10.1007/978-0-387-77516-6_22"
}

@article{Gowanlock2011,
author = {M.G. Gowanlock and D.R. Patton and S.M. McConnell},
title ="{A Model of Habitability Within the Milky Way Galaxy}",
journal = {Astrobiology},
volume = {11},
number = {9},
pages = {855-873},
year = {2011},
doi = {10.1089/ast.2010.0555},
URL = {https://journals.sagepub.com/doi/abs/10.1089/ast.2010.0555}
}

@article{Lineweaver2004,
author = {Charles H. Lineweaver  and Yeshe Fenner  and Brad K. Gibson },
title = "{The Galactic Habitable Zone and the Age Distribution of Complex Life in the Milky Way}",
journal = {Science},
volume = {303},
number = {5654},
pages = {59-62},
year = {2004},
doi = {10.1126/science.1092322},
URL = {https://www.science.org/doi/abs/10.1126/science.1092322}}

@article{Vukotic2016,
    author = {Vukoti\'{c}, B. and Steinhauser, D. and Martinez-Aviles, G. and \'{C}irkovi\'{c}, M. M. and Micic, M. and Schindler, S.},
    title = "{‘Grandeur in this view of life’: N-body simulation models of the Galactic habitable zone}",
    journal = {Monthly Notices of the Royal Astronomical Society},
    volume = {459},
    number = {4},
    pages = {3512-3524},
    year = {2016},
    month = {07},
    issn = {0035-8711},
    doi = {10.1093/mnras/stw829},
    url = {https://doi.org/10.1093/mnras/stw829},
}

@article{Gonzalez2001,
title = "{The Galactic Habitable Zone: Galactic Chemical Evolution}",
journal = {Icarus},
volume = {152},
number = {1},
pages = {185-200},
year = {2001},
issn = {0019-1035},
doi = {10.1006/icar.2001.6617},
url = {https://www.sciencedirect.com/science/article/pii/S0019103501966175},
author = {Guillermo Gonzalez and Donald Brownlee and Peter Ward},
}

@ARTICLE{Hyytia2012,
  author={Hyyti\"{a}, Esa and Virtamo, Jorma and Lassila, Pasi and Ott, J\"{o}rg},
  journal={IEEE Communications Letters}, 
  title="{Continuum Percolation Threshold for Permeable Aligned Cylinders and Opportunistic Networking}", 
  year={2012},
  volume={16},
  number={7},
  pages={1064-1067},
  doi={10.1109/LCOMM.2012.051512.120497}}

@INPROCEEDINGS{HyytiaOtt2013,
  author={Hyyti\"{a}, Esa and Ott, J\"{o}rg},
  booktitle={2013 Proceedings IEEE INFOCOM}, 
  title="{Criticality of large delay tolerant networks via directed continuum percolation in space-time}", 
  year={2013},
  volume={},
  number={},
  pages={320-324},
  doi={10.1109/INFCOM.2013.6566787}}

@book{Chiu2013,
  title = {Stochastic Geometry and its Applications},
  author = {Chiu, Sung Nok and Stoyan, Dietrich and Kendall, Wilfrid S. and Mecke, Joseph},
  edition = {3rd},
  publisher = {Wiley},
  address = {Chichester},
  year = {2013},
  doi = {10.1002/9781118658222}
}

@book{MeesterRoy1996, 
place={Cambridge}, 
series={Cambridge Tracts in Mathematics}, 
volume = {119},
title={Continuum Percolation}, 
publisher={Cambridge University Press}, 
author={Meester, Ronald and Roy, Rahul}, 
year={1996}, 
collection={Cambridge Tracts in Mathematics},
doi = {10.1017/CBO9780511895357}
}

@article{Forgan2019, 
title={Exoplanet transits as the foundation of an interstellar communications network}, 
volume={18}, 
DOI={10.1017/S1473550417000283}, 
number={3}, 
journal={International Journal of Astrobiology}, 
author={Forgan, Duncan H.}, 
year={2019}, 
pages={189-198}}

@article{Lares2020, 
title={Monte Carlo estimation of the probability of causal contacts between communicating civilizations}, 
volume={19}, 
DOI={10.1017/S147355042000018X}, 
number={5}, 
journal={International Journal of Astrobiology}, 
author={Lares, M. and Funes, J. G. and Gramajo, L.}, 
year={2020}, 
pages={393-405}}

\end{document}